%% file: lcdc.tex
\documentclass[conference]{IEEEtran}
\usepackage{cite}
\usepackage{amsmath,amssymb,amsfonts}
\usepackage{algorithmic}
\usepackage{graphicx}
\usepackage{textcomp}
\usepackage{xcolor}
\usepackage[hyphens]{url}

\usepackage{fancyhdr}
\usepackage{multirow}

\usepackage{array}
\usepackage{booktabs}
\usepackage{hhline}
\usepackage{tabularx}
\usepackage[per-mode=repeated-symbol,binary-units=true]{siunitx}
\DeclareSIUnit{\belmilliwatt}{Bm}
\DeclareSIUnit{\dBm}{\deci\belmilliwatt}
\usepackage{soul}
\usepackage{enumitem}

\usepackage{mathptmx} %
\usepackage[normalem]{ulem}
\usepackage{flushend}
\usepackage{dblfloatfix}
\usepackage{float}

\definecolor{maroon}{RGB}{128, 0, 0}
\usepackage{tikz}
\usepackage{xparse}
\NewDocumentCommand{\circledwrites}{
    O{maroon}
    O{black}
    O{white}
    m
    }{
    \tikz[baseline=(char.base)]{
    \node[shape=circle, fill=#1, draw=#2, text=#3, inner sep=0.5pt] (char) {#4};}}

\definecolor{orange}{RGB}{237, 125, 49}
\NewDocumentCommand{\circledreads}{
    O{orange}
    O{black}
    O{white}
    m
    }{
    \tikz[baseline=(char.base)]{
    \node[shape=circle, fill=#1, draw=#2, text=#3, inner sep=0.5pt] (char) {#4};}}

\definecolor{palegreen}{RGB}{112, 173, 71}
\NewDocumentCommand{\circledRAM}{
    O{palegreen}
    O{black}
    O{white}
    m
    }{
    \tikz[baseline=(char.base)]{
    \node[shape=circle, fill=#1, draw=#2, text=#3, inner sep=0.5pt] (char) {#4};}}

\NewDocumentCommand{\circleddefault}{
    O{gray}
    O{black}
    O{white}
    m
    }{
    \tikz[baseline=(char.base)]{
    \node[shape=circle, fill=#1, draw=#2, text=#3, inner sep=0.5pt] (char) {#4};}}

\newcommand*{\Lightning}[1][]{%
  \tikz[
    x=.55 * height("H"),  %
    y=height("H"),  %
    baseline = -3mm,
    line width=.02 * height("H"),
    line join=bevel,
  ]
  \filldraw[{#1}]
    (-.4, -.5) -- (.4, -.03) -- (-.1, .06) --
    (.4, .5) -- (-.4, .03) -- (.1, -.06) --
    cycle
    (-.5 - .08, 0)
    (.5 + .08, 0)
  ;%
}
\newcommand*{\LCDCTitle}[1][]{%
  LC{\Lightning[line join=round, xscale=1.6, yscale=1.6]}DC:
}

\newcommand*{\LightningText}[1][]{%
  \tikz[
    x=.55 * height("H"),  %
    y=height("H"),  %
    baseline=(current bounding box.south),
    line width=.02 * height("H"),
    line join=bevel,
  ]
  \filldraw[{#1}]
    (-.4, -.5) -- (.4, -.03) -- (-.1, .06) --
    (.4, .5) -- (-.4, .03) -- (.1, -.06) --
    cycle
    (-.5 - .08, 0)
    (.5 + .08, 0)
  ;%
}
\newcommand*{\LCDC}[1][]{%
  LC{\LightningText}DC
}

\usepackage{relsize}
\newcommand{\subtitlerelsize}{2} %
\newcommand{\subtitlelinesep}{0.2em} %

\makeatletter
\let\MYcaption\@makecaption
\makeatother

\usepackage[font=footnotesize]{caption, subcaption}

\makeatletter
\let\@makecaption\MYcaption
\makeatother

\DeclareSIUnit{\belmilliwatt}{Bm}
\DeclareSIUnit{\dBm}{\deci\belmilliwatt}

\newcolumntype{C}[1]{>{\centering\arraybackslash}p{#1}}
\newcolumntype{L}[1]{>{\raggedright\arraybackslash}p{#1}}

\def\BibTeX{{\rm B\kern-.05em{\sc i\kern-.025em b}\kern-.08em
    T\kern-.1667em\lower.7ex\hbox{E}\kern-.125emX}}

\usepackage[bookmarks=true, breaklinks=true, letterpaper=true, colorlinks, linkcolor=black, citecolor=black, urlcolor=black]{hyperref}

\title{\huge Energy-Proportional Data Center Network Architecture Through OS, Switch and Laser Co-design\\[\subtitlelinesep]%
    \smaller[\subtitlerelsize]{}Think Green - Turn Off The Lights}
\author{

\IEEEauthorblockN{
Haiyang Han\IEEEauthorrefmark{1},
Nikos Terzenidis\IEEEauthorrefmark{2},
Dimitris Syrivelis\IEEEauthorrefmark{3},
Arash F. Beldachi\IEEEauthorrefmark{4},
George T. Kanellos\IEEEauthorrefmark{4},
Yigit Demir\IEEEauthorrefmark{5},~~~\\
Jie Gu\IEEEauthorrefmark{1},
Srikanth Kandula\IEEEauthorrefmark{6},
Nikos Pleros\IEEEauthorrefmark{2},
Fabi\'an Bustamante\IEEEauthorrefmark{1} and
Nikos Hardavellas\IEEEauthorrefmark{1}
\vspace{-6pt}
}
\and

\IEEEauthorblockA{
~~\IEEEauthorrefmark{1}CS \& ECE, Northwestern University~~~\\
~~haiyang.han@u.northwestern.edu\\
~~\{jgu, fabianb, nikos\}@northwestern.edu
\vspace{-6pt}
}
\and

\IEEEauthorblockA{
~~\IEEEauthorrefmark{2}Aristotle University of Thessaloniki, Greece~~~~\\
~~\{nterzeni, npleros\}@csd.auth.gr
\vspace{-6pt}
}
\and

\IEEEauthorblockA{
~~\IEEEauthorrefmark{3}NVIDIA, Israel~~~~~\\
~~dimitriss@nvidia.com
\vspace{-6pt}
}
\and

\IEEEauthorblockA{
~~~~~~~\IEEEauthorrefmark{4}EEng, University of Bristol, UK~~\\
~~~~~~~\{gt.kanellos, Arash.Beldachi\}@bristol.ac.uk
\vspace{-20pt}
}
\and

\IEEEauthorblockA{
\IEEEauthorrefmark{5}Google~~\\
yigit@u.northwestern.edu
\vspace{-20pt}
}
\and

\IEEEauthorblockA{
\IEEEauthorrefmark{6}Microsoft Research~~\\
srikanth@microsoft.com
\vspace{-20pt}
}

}

\begin{document}

\bstctlcite{IEEEexample:BSTcontrol}

\maketitle
\thispagestyle{plain}
\pagestyle{plain}

\begin{abstract}
Optical interconnects are already the dominant technology in large-scale data center networks. However, the high optical loss of many optical components coupled with the low efficiency of laser sources result in high aggregate power requirements for the thousands of optical transceivers used by these networks. As optical interconnects stay always on even as traffic demands ebb and flow, most of this power is wasted. We present \LCDC, a data center network system architecture in which the operating system, the switch, and the optical components are co-designed to achieve energy proportionality.

\LCDC capitalizes on the path divergence of data center networks to turn on and off redundant paths according to traffic demand, while maintaining full connectivity. Turning off redundant paths allows the optical transceivers and their electronic drivers to power down and save energy. Maintaining full connectivity hides the laser turn-on delay.
At the node layer, intercepting send requests within the OS allows for the NIC's laser turn-on delay to be fully overlapped with TCP/IP packet processing, and thus egress links can remain powered off until needed with zero performance penalty.

We demonstrate the feasibility of \LCDC by i) implementing the necessary modifications in the Linux kernel and device drivers, ii) implementing a \SI{10}{\giga\bit\per\second} FPGA switch, and iii) performing physical experiments with optical devices and circuit simulations.
Our results on university data center traces and models of Facebook and Microsoft data center traffic show that \LCDC saves on average 60\% of the optical transceivers power (68\% max) at the cost of 6\% higher packet delay.
\end{abstract}

\input{introduction}
\input{motivation}
\input{architecture}

\input{feasibility}
\input{methodology}
\input{results}
\input{related}
\input{conclusion}

\section*{Acknowledgements}
This work was partially supported by NSF awards CCF-1453853 and CCF-1846424.

\bibliographystyle{IEEEtranS}
{\raggedbottom\bibliography{IEEEabrv.bib,local.bib}}

\end{document}

%% file: introduction.tex
\section{Introduction}
\label{intro}
Optical interconnects have emerged as a promising solution to meet the growing demand for high-bandwidth, low-latency, and energy-efficient communication in data centers~\cite{abts:energy-proportional-optical,greenberg:vl2,roy:facebook-network}. A significant fraction of the energy consumption in these networks can be attributed to the laser sources and laser drivers. As we argue, most of this energy is wasted.

For example, a typical 64-port switch with \SI{220}{\watt} peak power draws on average \SI{140}{\watt}~\cite{arista:7050} and employs 10G SFP+ optical transceivers at \SI{1}{\watt} per port. With this configuration the switch consumes a third of its power on optical transceivers. QSFP (40G) transceivers consume as much as 3.5$\times$ more power, raising the power ceiling even further.
However, most of the laser energy is wasted. Unlike traditional interconnects that expend most of their energy only during packet transmission, optical interconnects are always on and consume power, even during periods of inactivity. In reality, the interconnect often stays idle for long periods: compute-intensive workloads underutilize the interconnect (common in scientific and many analytic workloads), and servers in data centers often stay idle or exhibit load imbalances (data centers are typically 20--30\% utilized~\cite{barroso:energy-proportionality, barroso2013datacenter}).

The natural solution is to turn off the transceiver of an idle link to save energy, and turn it back on when packets arrive to facilitate communication~\cite{heller:elastic-tree}. A naive implementation of this ``\textit{laser gating}'', though, risks exposing multiple laser turn-on delays to the end application as a packet typically crosses multiple links to reach its destination, significantly increasing packet latency and lowering performance.

We propose to hide the laser turn-on delay by capitalizing on the path diversity of modern data center interconnects. To service high levels of traffic across a large number of nodes, data centers typically exploit scalable network topologies. For example, full-optical Clos networks are widely deployed in Facebook~\cite{roy:facebook-network}, Google~\cite{singh:jupiter} and Microsoft~\cite{greenberg:vl2} data centers, and flattened butterfly topologies have been proposed as a cost-efficient alternative by Google~\cite{abts:energy-proportional-optical}. All these topologies provide path diversity, i.e., there are multiple paths between any source-destination pair. Instead of turning off links arbitrarily and severing end-to-end paths, which exposes the laser turn-on delay, we propose to turn off only redundant links when utilization is low, and turn them on again when the aggregate workload needs more bandwidth. Maintaining full connectivity removes the laser turn-on latency from the critical path and results in minimal performance degradation.

We also propose to control the server-to-ToR (Top of the Rack) switch links by intercepting socket write calls at the OS level and raising a signal to the Network Interface Card (NIC) lasers to turn on. The TCP/IP processing latency is high enough that allows for ample time to notify the server NIC of the impending traffic. By the time the data are ready to be sent, the laser is already on and locked at the appropriate frequency, thereby allowing the node to save energy with zero performance penalty.

The main idea of laser gating is not new. Such techniques have been proposed before for on-chip interconnects~\cite{demir:ecolaser, demir:ecolaser-plus, demir:lac, demir:slac}. An earlier study capitalizing on path diversity for laser control in on-chip interconnects~\cite{demir:slac} was extended to data centers but only as a conceptual study; no evaluation was performed on traffic similar to modern large-scale data centers (only a university data center traffic trace was used), and the feasibility of the technique was not proven or discussed. After all, the latency to perform network control plane changes has been previously shown to be in the \SI{}{\milli\second} scale~\cite{farrington:helios}, which could render such a technique useless, and the latency to perform link retraining for Clock and Data Recovery (CDR) can be prohibitively high.

In this paper we aim to set the record straight. We evaluate laser gating on multi-stage data center networks with a Clos architecture similar to the one found in modern hyperscale data centers~\cite{greenberg:vl2, roy:facebook-network, singh:jupiter} using traffic patterns that closely approximate real-world traffic~\cite{greenberg:vl2, kandula:traffic-nature, roy:facebook-network}. We demonstrate that control plane changes can be performed in a matter of \SI{}{\nano\second} by implementing \LCDC on a $6 \times 6$ \SI{10.8}{\giga\bit\per\second} switch on an FPGA. We implement a device driver on a modern Linux operating system and kernel changes that intercept socket write calls to raise a signal and alert the NIC of imminent outgoing traffic. We measure the latency of the TCP/IP processing and show that the NIC has ample time to turn on its transceiver while the outgoing packet is being prepared. Finally, through a combination of physical experiments and analog SPICE simulations of laser driver circuit models, we show that optical transceivers and their electronic drivers can be turned on at \SI{}{\micro\second} scales. Collectively, these results demonstrate that laser power gating at the data center scale is indeed feasible, and can be driven by the OS and network switch layers. We then estimate the energy savings that can be achieved on a variety of scenarios through simulations and technology projections.

More specifically, our contributions are:

\begin{itemize}
    \item{We propose \LCDC (Laser Control for Data Centers), a data center network system architecture in which the operating system, the switch, and the optical components are co-designed to achieve energy proportionality.}
    \item{We demonstrate the feasibility of employing \LCDC at servers and global switches through modifications in the Linux kernel and device drivers, switch design on FPGA boards, physical experiments with optical devices, and analog circuit simulations.}
    \item{We develop a data center traffic generator that models the traffic exhibited at Facebook and Microsoft. We show that our traffic generator produces CDFs of flow size and flow intervals that closely match real-world traffic.}
    \item{We evaluate \LCDC on models of Facebook and Microsoft data center traffic, as well as traffic traces from a university data center. \LCDC saves on average 60\% of the optical transceiver power (68\% max) at the cost of 6\% higher packet delay.}
    \item{As servers and cooling, electrical and mechanical systems become increasingly more energy efficient, we project that the network will command a larger fraction of the overall data center energy consumption. We estimate the potential savings of \LCDC on a hypothetical future datacenter that applies multiple server-level energy optimizations. We find that \LCDC can save 12\% and 21\% of the data center energy on average when deactivating transceivers or transceivers and switch PHY and NIC electronics, respectively, even for cases where the the server utilization approaches 70\%.}
\end{itemize}

%% file: motivation.tex
\section{Motivation}
\label{motivation}
Power and energy efficiency have been at the forefront of research in circuits and computer architecture for at least 15 years~\cite{falsafi:pacs-2000, stan:power-aware-computing, brooks:wattch}. Innovations in these fields coupled with technological advances in materials, semiconductor processes and packaging yield lower-power devices at each technology node~\cite{sia:irds2020, toshiba:16-die-nand, borkar:ipdps13-keynote, kultursay:sttram}. The end result is the rapidly increasing energy efficiency of server components, including memory, storage, processors, and ultimately the servers themselves. Simultaneously, the high power demands of modern data centers has pushed data center operators to drastically reduce power inefficiencies. As a result, modern state-of-the-art data centers reduced the power overhead for cooling, electrical and mechanical systems from more than 2$\times$ a decade ago to only 6\% today~\cite{facebook:prineville-pue, google:pue, meta:pue-board}. As servers deliver increasingly higher performance per Watt, and data center overheads are aggressively eliminated, their contribution to the overall data center power consumption drops, exposing other components that have not yet received similar attention~\cite{heller:elastic-tree}. Data center networks are one of these components, and its relative power consumption rises as innovations in other sectors reduce the power draw of other data center components.

\begin{figure*}[!ht]
    \centering
    \includegraphics*[width=\textwidth]{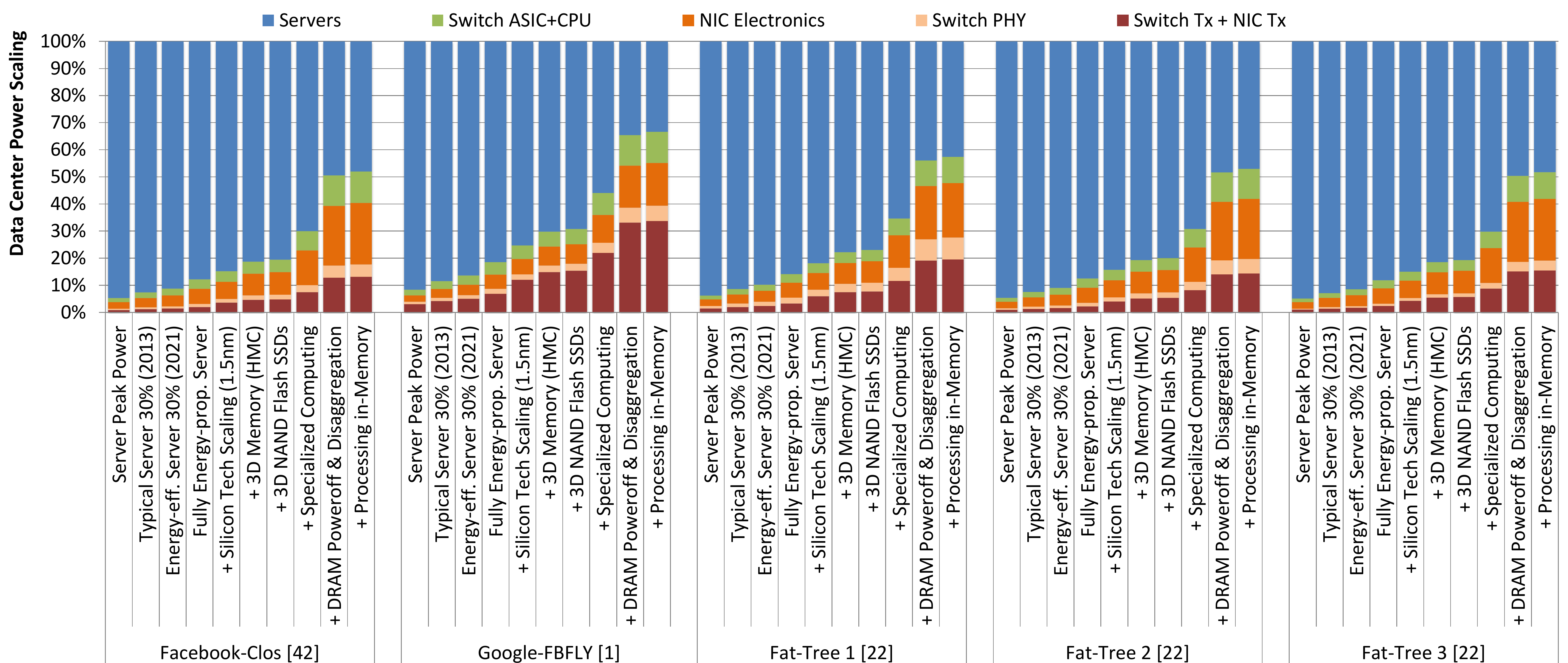}
    \caption{\textbf{Data center power breakdown change as a function of server system optimizations, across various network designs.}}
    \label{fig:power}
\end{figure*}

Figure~\ref{fig:power} shows the breakdown of data center power as various optimizations are applied on servers. For generality, we carry out the same study across various network designs from the literature: a Clos Facebook site~\cite{roy:facebook-network}, a Flattened Butterfly interconnect by Google~\cite{abts:energy-proportional-optical}, and three Fat-Tree networks derived from~\cite{farrington:merchant-silicon-hoti} that are either readily available using off-the-shelf components (Fat-Tree 1), or require board and chassis engineering for higher efficiency and lower cost (Fat-Tree 2), or require board, chassis and new ASIC design (Fat-Tree 3). All designs are modeled using the exact component counts and connectivity described in their respective publications.
Figure~\ref{fig:power} breaks down the data center power into:

\begin{itemize}
    \item{Server systems}
    \item{Switch ASIC and CPU chips (\SI{28}{\watt} per switch)~\cite{farrington:merchant-silicon-hoti}}
    \item{Server network card (NIC) electronics (\SI{10}{\watt})~\cite{abts:energy-proportional-optical}}
    \item{Switch PHY chips that implement the Layer 1 protocol exposed by the switch (\SI{0.8}{\watt}, one PHY per port)~\cite{farrington:merchant-silicon-hoti}}
    \item{Optical transceivers on switch and NIC card ports, assuming \SI{1}{\watt} for 10G SFP+, \SI{2.4}{\watt} for 40G QSFP~\cite{farrington:merchant-silicon-hoti}}
\end{itemize}

The figure excludes the overhead for cooling, electrical and mechanical systems within the data center, as the additional overhead these systems impose is proportional to the power draw of the server, storage and networking equipment. We note that modern hyperscale data centers often achieve Power Usage Effectiveness (PUE) of 1.06~\cite{google:pue, facebook:prineville-pue, meta:pue-board} and report a comprehensive trailing twelve-month PUE of 1.10~\cite{google:pue} across all large-scale data centers, in all seasons, including all sources of overhead. Thus, the additional overhead of large-scale data centers nowadays represents a relatively small fraction of their overall energy consumption.

Most readers would be familiar with the left-most stacked bar for each network, where 92--95\% of the power is consumed by the servers and only 5--8\% goes to the interconnect. However, this assumes that all servers run at 100\% utilization and draw peak power all the time. In reality, servers are typically only 20--30\% utilized~\cite{barroso:energy-proportionality, barroso2013datacenter}. Thus, their contribution to data center power is lower, and the relative importance of other components grows. The second stacked bar shows this effect assuming typical servers circa 2013~\cite{fan:power-provisioning}.

The third bar shows the effect of 30\% server utilization on a best-of-class modern server that is nearly energy-proportional: the Lenovo Think System SR665, which currently has the highest performance per Watt as measured by an audited SPECpower benchmark~\cite{specpower:lenovo-2020}. This server expends 58\% of its power at 30\% utilization (compared to 70\% for the 2013 server~\cite{barroso2013datacenter}). The fourth bar continues that trend and models a fully energy-proportional server at 30\% utilization~\cite{barroso:energy-proportionality, fan:power-provisioning, barroso2013datacenter} at which point it consumes 40\% of its peak power. On average, a data center with fully energy-proportional servers at 30\% utilization seems to spend about 86\% of its energy on the servers and the remaining 14\% on the network.

The following bars apply successively more optimizations on the server components (and if applicable the switch and NIC electronics) to account for the transition from \SI{7}{\nano\meter} to \SI{1.5}{\nano\meter} CMOS technology following IRDS projections~\cite{sia:irds2020, borkar:isc15-keynote}; the use of modern memory technology (e.g., Micron's 3D hybrid memory cube, HMC)~\cite{pawlowski:hmc, borkar:isc15-keynote}; the introduction of 16-die-stacked 3D NAND Flash for solid-state drives~\cite{toshiba:16-die-nand, andersen:rethinking-flash}; and the employment of specialized computing which is modeled after the Catapult project~\cite{putnam:catapult} in Microsoft that deployed FPGAs in production data centers to off-load computations from conventional general-purpose processors.

Finally, the last two bars show the impact of applying various recent technologies such as DRAM refresh reduction~\cite{kim2020charge}, DRAM idle power-off~\cite{zhang2014dimmer}, memory disaggregation~\cite{nitu2018welcome}, and near-memory processing~\cite{ke2021near}. We model the impact of these optimizations by employing them only on the appropriate components, following the typical power profile of data-center-class servers~\cite{fan:power-provisioning} and switches~\cite{arista:7050, farrington:merchant-silicon-hoti}.

As Figure~\ref{fig:power} indicates, with each optimization the relative power consumption of servers drops, to the point where the network becomes a major component. Our projections indicate that, unless something is done, the thousands of transceivers employed in a data center network will account for 20\% of the data center power consumption on average across network designs, after these series of optimizations are applied on the servers. Moreover, the combined switch PHY chips, server NIC electronics and transceivers will account for as much as 46\% of the data center power. Thus, we argue it is time to start optimizing the network not only for latency, bandwidth and cost, but also for power and energy efficiency.

\begin{figure*}[t]
    \centering
    \includegraphics[width=1.3\columnwidth]{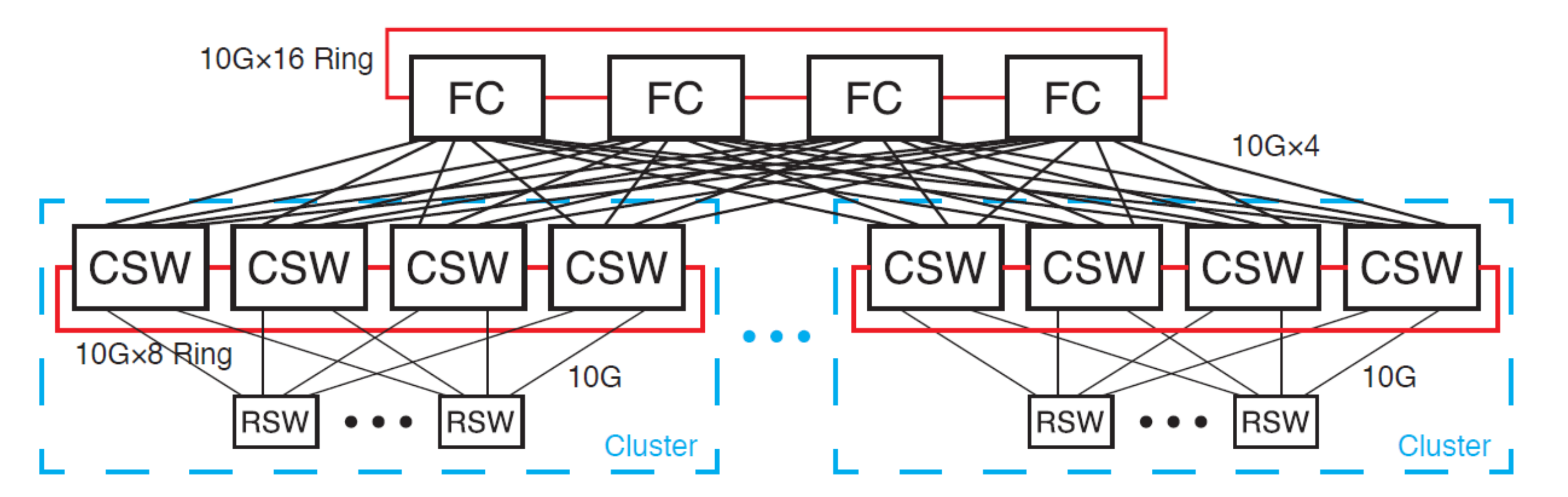}
    \caption{\textbf{Data center network configuration, similar to~\cite{roy:facebook-network}.}}
    \label{fig:config}
\end{figure*}

It is important to note that the optimizations we model in this study are carefully selected to be conservative and readily available, and they are solidly backed by experimental characterization of existing products. The technology projections conform to the roadmap that the semiconductor industry collectively sets every year since 1993~\cite{sia:irds2020, borkar:isc15-keynote}, the memory and SSD devices are already commercially available~\cite{pawlowski:hmc, nvidia:p100, computerworld:hmc, toshiba:16-die-nand}, nearly full-energy-proportional servers are commercially available today~\cite{specpower:lenovo-2020}, and some data centers already deploy specialized computing (e.g., Microsoft's Catapult~\cite{putnam:catapult}, Google's Tensor Processing Units~\cite{google:tensor-processor}).

To capture a wider range or projections, we also include two more sophisticated energy-efficiency optimizations that have been in active development in industry for more than a decade, and while they are not mainstream products yet, they are in advanced stages of development. These optimizations include near-memory processing~\cite{ke2021near} and disaggregation~\cite{nitu2018welcome}.
There is a large number of much more aggressive optimizations that we explicitly chose not to include, as their ability to scale up to production at reasonable cost is unknown, or they are not a good fit for hypercale data centers, or simply because they are not commercially available yet, despite their high potential (e.g., STT-RAM, PCM, near-threshold-voltage processors, spintronics, neuromorphic processors, and chip- and board-level photonics).

%% file: architecture.tex
\section{\LCDC Architecture}
\label{arch}
We propose \LCDC to minimize the amount of power spent on optical transceivers employed in switches and server NIC cards by turning them off when they are not needed. Moreover, we can extend the design of \LCDC to also put the switch PHY chips and the server NIC electronics into a low-power state at the same time that the laser is switched off. While we do not study this extension in this paper, it can address as much as 46\% of the projected data center power consumption (Figure~\ref{fig:power}), even after accounting for the CMOS scaling of the PHY and NIC electronics.

To minimize the performance impact of waiting for the lasers to turn on, \LCDC deactivates only redundant links, while maintaining full network connectivity. As there is always a path connecting any pair of nodes, packets can still reach their destination while links are activated. \LCDC monitors the interconnect traffic and turns off links when the utilization is low to save energy, and activates additional links when the utilization is high to increase performance.

\subsection{\LCDC Operation at the Switch Level}
\label{switch}
\LCDC is applicable to any network topology with path diversity. In this paper we evaluate it in a network similar to a site in the Facebook data center~\cite{roy:facebook-network}. Figure~\ref{fig:config} presents the network design. Each rack has 48 nodes which connect to a Rack Switch (RSW). 32 RSW's form an Cluster and connect to 4 Cluster Switches (CSW). 4 Clusters (16 CSWs) connect to 4 ``Fat Cat'' Routers (FC) to form a site. The RSWs have 48 10G (downlink) input-output ports and 4 10G uplinks to CSW intermediate routers (12:1 oversubscription). In turn, the CSW's provide 4 40G uplinks to FC switches (2:1 oversubscription). Each set of CSWs within a cluster and the FCs are connected on a ring formed by 8 10G links and sixteen 10G links respectively, for load balancing.

\LCDC controls each tier independently. Each RSW has 4 uplinks to connect to all the CSWs in its cluster (path divergence). Each one of these uplinks defines a \textit{Stage}. When we say that stage k is active, we mean that links $1$ through $k$ are active. Initially only one stage per RSW is active. \LCDC on each switch estimates network traffic by monitoring the buffer depth (buffer utilization, aka queue backlog) of its active links, which is an accurate and lightweight method~\cite{chen:mp3}. When a buffer's depth exceeds a tunable threshold (high watermark), the RSW turns on an additional stage to provide higher bandwidth and minimize queueing delay. The switch sends a control message through the already active stages to the corresponding CSW informing it of the new stage activation, and once its transceiver is active and has received an acknowledgement from the CSW that the receiving side is active, it starts using the additional link.

The newly activated stage turns off when the RSW that activated it becomes underutilized, i.e., its buffer utilization falls below a low watermark. In that case, \LCDC realizes that the additional bandwidth provided by the redundant link is not necessary, and should turn it off. The RSW stops receiving outgoing messages in this port, serves all the packets in its buffers, and then notifies the corresponding CSW with a stage turn-off message which deactivates the last activated stage. Once the CSW acknowledges the link deactivation, RSW turns off its transceiver too. Link activation and deactivation at CSWs and FCs is performed similarly.

\begin{figure*}[t]
    \centering
    \includegraphics[width=1.3\columnwidth]{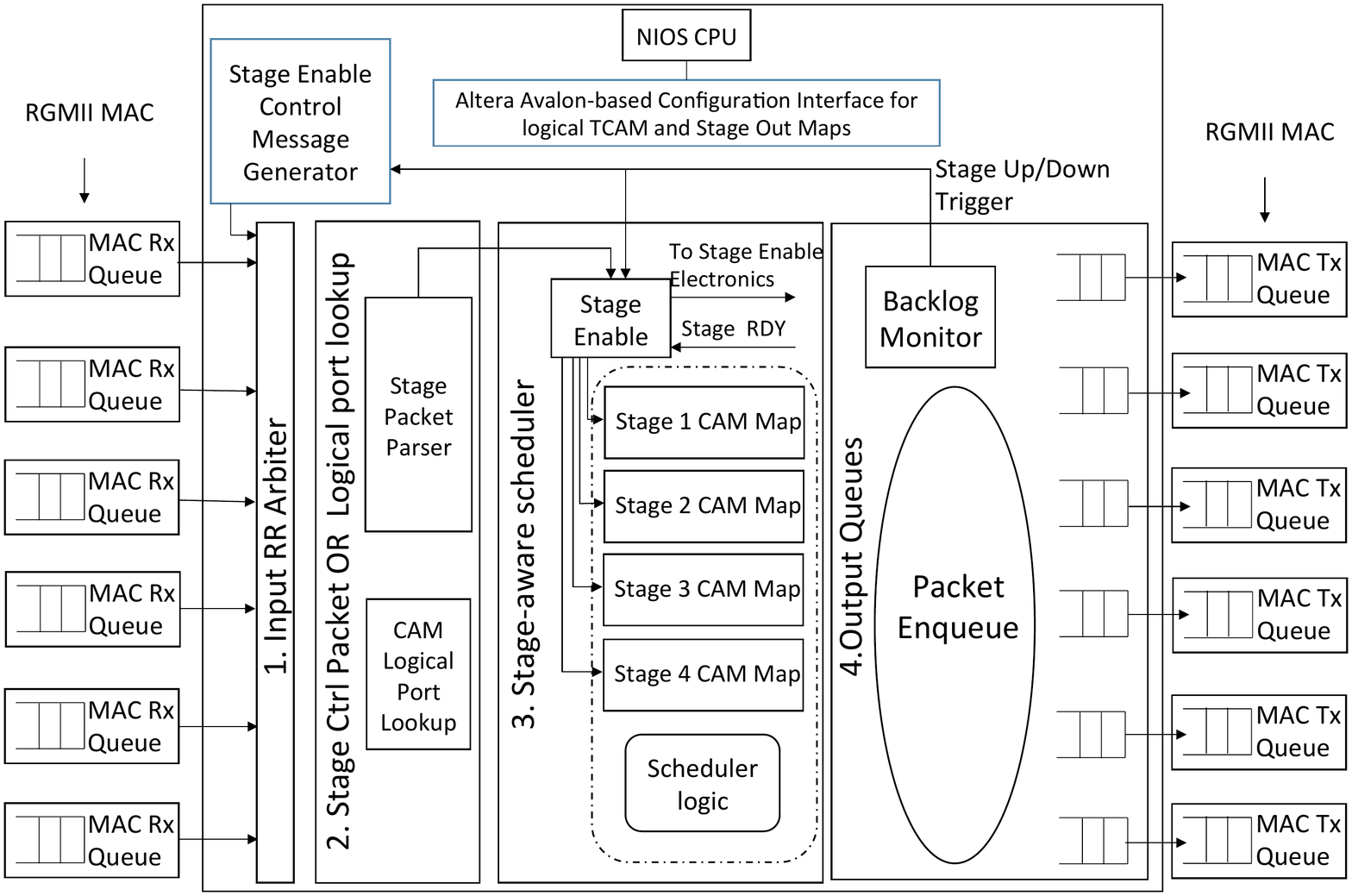}
    \caption{\textbf{\LCDC switch architecture.}}
    \label{fig:arch}
\end{figure*}

\LCDC adaptively routes traffic through only the active stages which achieves load balancing, turns on lasers on the side and hides their turn-on delay, and avoids unnecessary link activation which maximizes the energy savings.

\subsection{\LCDC Switch Architecture}
\label{switcharch}
Figure~\ref{fig:arch} shows the architecture of the \LCDC switch. \LCDC is a combined input-output queued switch (CIOQ)~\cite{shang-tse:output-queueing}. At the input, buffering relies on the hardware queues of RGMII MAC, while at the output the design features one RAM-based queue per physical port. The \LCDC datapath is 64-bit wide and adds a single annotation flit to each packet that flows through it to pass information from one stage to the next. The design implements a control message channel that uses the same physical ports as data packets (in-band).

\subsubsection{Dataplane Pipeline Stages}
\label{pipeline}

The Ethernet frames are pulled from RGMII MAC queues that drive the physical interfaces using a round robin arbiter. Besides the physical input ports, the switch features a virtual port that interfaces a 2-port memory to the arbiter. The memory is pre-programmed with all flavors of control packets that are forwarded to other switches to initiate \LCDC stage changes. The arbiter polls the virtual port out-of-order to prioritize the forwarding of generated control packets.

In the next pipeline stage the processing differs depending on whether the packet is \LCDC control or not. The \LCDC control packet is an ethernet frame where bytes 13-14 form the \LCDC Ethernet type (0x9100). The next 8 bytes contain the \textit{senderID} (\SI{4}{\byte}), the \textit{stageID} (\SI{2}{\byte}), and the \textit{TTL} (\SI{2}{\byte}). The packet is then padded to the minimum ethernet frame size. The \textit{stageID} denotes the \LCDC stage to be enabled and the possible values are deployment-wide agreed. The \textit{TTL} designates the number of switching layers that the control packet may flow through before it is discarded. This approach simplifies the distribution scheme by allowing control packets to get forwarded from all output ports of each switch without worrying for endless loops. Accordingly, the control packet processing checks the \textit{stageID} and sends the appropriate notification to the \LCDC stage enabling component, updates the \textit{TTL} and drops the packet (if zero) or propagates it for forwarding. Finally, the \textit{senderID} designates the control packet sender, so it is used to determine if the packet was generated in the local switch. If it was, it is directly forwarded to the scheduler because the stage transition process is initiated before control packet generation.

The non-local packets initiate a Content Addressable Memory (CAM) lookup to match the packet destination Ethernet MAC address with a logical port, which is a deployment-unique identifier of the destination switch where the packet recipient is attached. This switch addressing scheme is internal to \LCDC and has to be configured by the control plane (Section~\ref{control}). For multicast support, special logical port identifiers are programmed by the control plane for each multicast tree. Accordingly, the logical port identifier is pushed to the packet annotation space and a multicast bit is set when needed before the packet propagates to the scheduler pipeline stage.

\begin{figure*}[!ht]
    \centering
    \begin{subfigure}{0.4\textwidth}
       \centering
       \includegraphics[width=\textwidth]{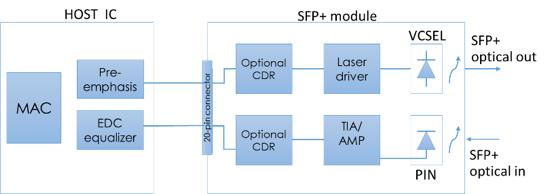} \\[\abovecaptionskip]
       \caption{SFP+ transceiver module configuration.}
       \label{fig:sfp_config}
    \end{subfigure}
    \hfill
    \begin{subfigure}{0.36\textwidth}
       \centering
       \includegraphics[width=\textwidth]{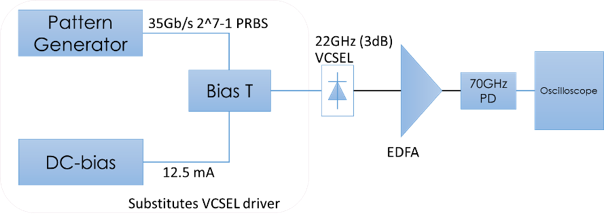} \\[\abovecaptionskip]
       \caption{Experimental setup for the optical modules.}
       \label{fig:sfp_setup}
    \end{subfigure}
    \hfill
    \begin{subfigure}{0.18\textwidth}
       \centering
       \includegraphics[width=\textwidth]{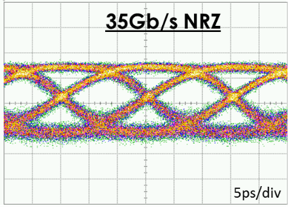} \\[\abovecaptionskip]
       \caption{Output Eye diagram.}
       \label{fig:sfp_eye}
    \end{subfigure}

    \caption{\textbf{Physical on/off timing experiment with a VCSEL laser device.}}
    \label{fig:sfp}
\end{figure*}

The scheduler load-balances the incoming traffic over the available output physical port queues based on the packet destination and the \LCDC stage that is enabled. As the provided packet destination is a logical port, the scheduler has access to a series of binary CAM tables, one per \LCDC stage, to determine the physical output port options. These CAM tables take the logical port as input and provide a binary map at the output where all the possible physical ports that may be used for this destination, based on the enabled \LCDC stage, are one-hot encoded. The scheduler uses each time a single CAM that corresponds to the currently enabled stage. A simple weighted scheduling algorithm chooses the output queue with the minimum backlog from the ones that belong to the given map. In case of multicast packets, the scheduler places a copy to all output queues of the encoded map, which is encoded accordingly to implement the multicast. This step concludes the \LCDC datapath operation.

Moreover, three peripheral components orchestrate the \LCDC stage transitions. First, the queue backlog monitor component inspects all the output queue backlogs. When a backlog exceeds an administrator defined high watermark, the stage up trigger is sent in parallel to the other two components: i) the stage enable component and ii) the stage enable control message generator. The former immediately drives the electronics to enable the next stage and when it receives a stage ready signal it enables the CAM stage table that corresponds to the currently active output ports. The latter activates the virtual input port so that the arbiter pulls the proper control packet to notify the rest of the deployment for the state change. If the backlogs become low, the backlog monitor sends to the stage enable component the stage down trigger, which immediately enables the appropriate CAM table while it requests the shutdown of specific physical ports.

\subsubsection{\LCDC Control Plane}
\label{control}
The \LCDC switch requires control plane support that has overview of the whole network deployment and device topology in order to provide the required forwarding information on all CAM tables, the thresholds on the backlog monitors and the programming of the control packets for the underlying switch fabrics. For this particular purpose the current design features an Altera Avalon bus interface that connects the aforementioned components to the Altera NIOS processor, a softcore processor solution that is provided by the FPGA platform we use. Therefore, low-level software-based control plane support can be realized on each switch, which can then be integrated with data center network orchestration tools like OpenStack Neutron according to the SDN paradigm.

\subsection{OS and Device Driver Design for Node-Level \LCDC}
\label{os}
In addition to controlling the redundant links, \LCDC independently controls the transceivers on each server's NIC card, by using a modified network card device driver which is controlled by the system as a kernel module. \LCDC intercepts the Linux \texttt{sendmsg()} system call and replaces it with our version of \texttt{sendmsg()} by modifying the system call table at driver module initialization. Upon a user-level \texttt{socket{\_}write()} call, the driver signals its laser to turn on and then invokes the original \texttt{sendmsg()} function. While the laser turns on, the payload goes through the TCP/IP stack processing. By the time the driver-specific transmit function is called to transmit over the fiber, sufficient time has elapsed for the laser to be fully operational and transmit the data. Changes to the Linux driver and kernel module is minimal at around 200 lines of code.

%% file: feasibility.tex
\section{Feasibility Study}
\label{feasibility}

\subsection{Feasibility at the Device Level}

\subsubsection{Optical Transceiver Turn-on Delay}
\label{transceiver}

The datacom transceiver modules that are largely deployed in current data center implementations mainly rely on SFP+ modules for up to \SI{10}{\giga\bit\per\second} link bandwidths and QSFP modules, which is a quadruple form of SFP+, for link bandwidths in the range of \SI{40}{\giga\bit\per\second} to \SI{100}{\giga\bit\per\second}. The simplest SFP+ form of transceiver is shown in Figure~\ref{fig:sfp_config} and comprises a transmitter side which includes a laser component followed by the laser driver circuitry, and a receiver side that includes a photodetector followed by the pre-amplifier, also called trans-impedance amplifier (TIA), and the post-amplifier. Optional additional circuitry may include a CDR (clock and data recovery) circuit in both transceiver sides.

In such transceiver configurations, the transmitter turn on/off timings are denoted as Tx\_Disable assert time / Tx\_negate assert time and are defined in the SFP+ multisource agreement (MSA)~\cite{sff:8431}. The MSA specifies that the transmitter turn on/off timings are \SI{100}{\micro\second} and \SI{1}{\milli\second} respectively, while the receiver turn on/off times denoted as Rx\_LOS assert delay / Rx\_LOS negate delay are \SI{100}{\micro\second} each. However, the MSAs have set the timing boundaries well beyond any safety margins of stable operation in order to fully relax the design and cost requirements of the transceivers~\cite{sff:8431}. In fact, though both optical and electrical components comprising the transceivers allow for significantly faster operations, this has never been a design objective for commercial manufacturers of Datacom transceiver modules, and hence they are not optimized for it.

This becomes evident for transceivers in the SFP+ form-factor that are deployed in PON (Passive Optical Network) applications, such as the 10GE-PON SFP+ transceivers. For these transceivers, the necessity for burst operation has led to their commercial implementations exhibiting turn on/off times of \SI{512}{\nano\second} each~\cite{eoptolink:eolx, igawa:10g-epon, delta:tep}. These transceivers retain all other Datacom SFP+ specifications, such as power consumption, bit rate, etc, and thus demonstrate that transceiver for Datacom applications (e.g., data centers) can also be implemented with equally fast \si{\micro\second}-scale turn on/off timings.

To provide further evidence that the optical components can be turned on/off at such high speed, we set an experimental setup of a manufactured VCSEL device and a simulation model of the electrical circuitry to assess the minimum timings required for turning on/off such transceivers.
Figure~\ref{fig:sfp_setup} shows the experimental setup that reveals the turn on/off timings for the optical components. We have chosen the \SI{22}{\giga\hertz} VCSEL laser of~\cite{kaur:2015flip} as the optical source, since VCSELs are typical laser sources for Datacom transceivers. We consider the non-return-to-zero (NRZ) direct modulation bandwidth as the laser turn on/off frequency. A pseudo-random binary sequence with a word length of $2^{7}-1$ and a bit rate of \SI{35}{\giga\bit\per\second} was generated from a commercial pattern generator. The output of the pattern generator was a single-ended NRZ signal with a swing of \SI{650}{\milli\volt} peak-to-peak, while a bias tee superimposed this data signal on a DC bias current of \SI{12.4}{\milli\ampere}. This setup substitutes the laser driving circuitry of Figure~\ref{fig:sfp_config}, and defines the electrical signal specs to be applied to the VCSEL through RF probes. The optical signal generated by the VCSEL was then pre-amplified through an erbium-doped fiber amplifier (EDFA) and launched to a Finisar XPDV3120R \SI{70}{\giga\hertz} photodetector with \SI{3}{\volt} bias. The electrical signal at the output of the PD was then captured by an Infinium sampling scope, revealing a clear eye-pattern for up to \SI{35}{\giga\bit\per\second} as shown in Figure~\ref{fig:sfp_eye}. This implies that both transmitter and receiver optical components have the ability to be turned on/off at timing below \SI{15}{\pico\second} respectively.

The experiment above demonstrates that the turn on/off speed achievable by laser devices is well below the \si{\micro\second}-scale that \LCDC requires. The electrical integrated circuit for the receiver can also be sufficiently fast for \LCDC application. An electrical circuit that even includes a burst mode CDR has been recently shown to exhibit optical power calibration in \SI{12.5}{\pico\second} and phase lock in \SI{18.5}{\pico\second}~\cite{rylyakov:201525}, both well below \SI{1}{\micro\second}.
We emphasize that switching off links and their lasers does not modify the links; the link characteristics are the same when the link is powered up again. This makes the link amenable to clock phase caching, which was recently demonstrated to provide clock and data recovery times below \SI{625}{\pico\second}~\cite{clark:cdr_nature, clark:cdr_ecoc, balani:sirius} on a real-time prototype with commercial transceivers, comfortably within the \si{\micro\second} requirement of \LCDC. This result was further validated against temperature variation and clock jitter~\cite{clark:cdr_nature}, demonstrating its resilience and applicability to real-world scenarios.

To fully assess the lower possible boundaries of the transceiver timing specifications, the only remaining component in our device-level feasibility study, we derived a SPICE-based analog simulation model for the laser driver depicted in Figure~\ref{fig:spice_model}. In this configuration, we considered the RLC electrical circuit equivalent for the VCSEL as provided by Finisar in~\cite{gazula:2010emerging}.

\subsubsection{Electronic Circuit Simulation}
\label{spice}

\begin{figure}[t]
    \centering
    \begin{subfigure}{0.215\textwidth}
       \centering
       \includegraphics[width=\textwidth]{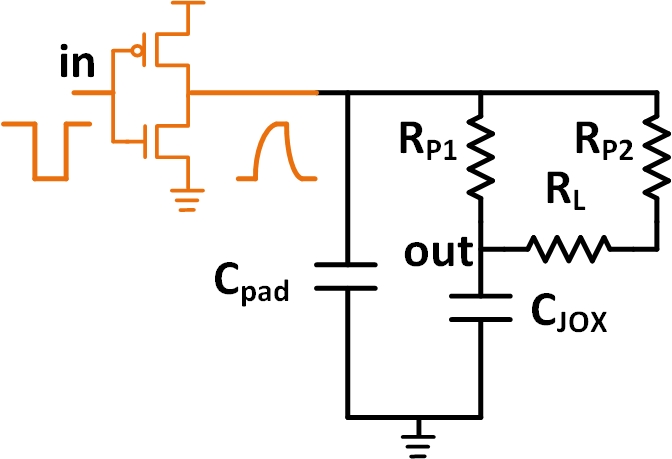} \\[\abovecaptionskip]
       \caption{Lumped model for VCSEL laser~\cite{gazula:2010emerging}.}
       \label{fig:spice_model}
    \end{subfigure}
    \hfill
    \begin{subfigure}{0.26\textwidth}
       \centering
       \includegraphics[width=\textwidth]{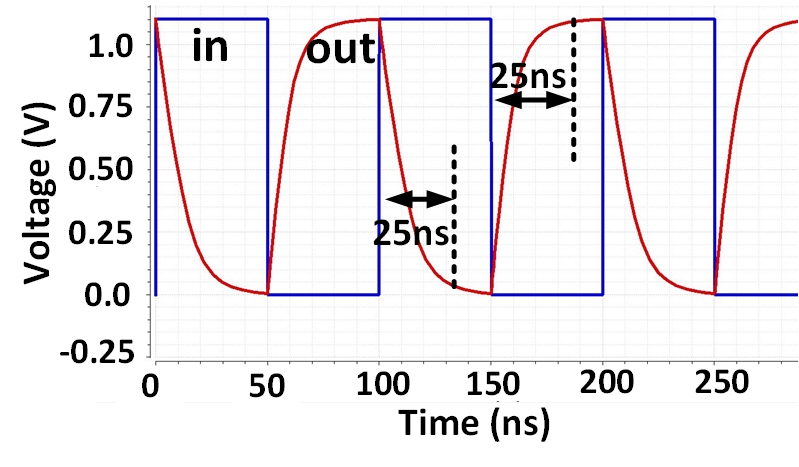} \\[\abovecaptionskip]
       \caption{Spice simulation waveform in a \SI{45}{\nano\meter} CMOS technology.}
       \label{fig:spice_waveform}
    \end{subfigure}

    \caption{\textbf{Spice simulation on power gating a VCSEL laser.}}
    \label{fig:spice}
\end{figure}

We created a spice simulation in a \SI{45}{\nano\meter} CMOS technology following the specification provided from VCSEL laser model~\cite{gazula:2010emerging}. As shown in Figure~\ref{fig:spice_model}, the model contains simple lumped components for bonding pad $C_{PAD}$ (\SI{68.3}{\pico\farad}), distributed resistance components $R_{P1} (215{\Omega})$, $R_{P2} (147{\Omega})$, and $R_{L} (1690{\Omega})$ for mirror, and combined junction and oxide capacitance $C_{JOX} (\SI{34}{\pico\farad})$. This simplified lumped model has been verified from DC to high frequency (\SI{25}{\giga\hertz}) to match well with measured responses from the physical VCSEL laser~\cite{gazula:2010emerging}. In our simulation, a small CMOS driver with transistor sizes of \SI{10}{\micro\meter} was used to turn on and off the laser block. Simulation in Figure~\ref{fig:spice_waveform} shows that the laser junction voltage can be driven within \SI{25}{\nano\second} with a large signal magnitude using our small CMOS driver. This simulation shows that such a laser source can be power gated well within \SI{100}{\nano\second} or \SI{1}{\micro\second} time window.

\subsection{Feasibility at the Switch Level}
\label{switch_feasibility}
Similar to transceiver modules, commercially available switches do not allow for fast nanosecond-scale updates to port maps or rules. While technologically this should be feasible within a few processor cycles, commercial designs never had the incentive to optimize it to a few nanoseconds.

To demonstrate the feasibility of \LCDC at the switch layer we implemented a prototype of the described \LCDC pipeline architecture in Verilog as a 6$\times$6 switch, with small-sized CAMs (100 entries each) and support for four \LCDC stages (Figure~\ref{fig:arch}). The design targets an Altera Stratix V GT platform where it achieves a clock rate of \SI{169.32}{\mega\hertz}, providing a \SI{10.8}{\giga\bit\per\second} backplane. The overall latency from the time a packet flit enters the pipeline until it is delivered to the output queues is 7 cycles (2 cycles for the logical port lookup, 2 for the stage out port map lookup, 2 for the scheduler, and 1 for placing at the output queue). The cycle count is expected to grow with the number of output ports because the scheduler checks all backlogs before queueing a packet.

Our design also demonstrates fast stage trigger generation. When the backlog monitor observes a threshold violation (checked on every cycle) it signals the stage enable component in the same cycle ($<$\SI{5.8}{\nano\second} delay). When a control packet that initiates the stage change appears at the input, 2 cycles elapse before the proper flit is parsed (\SI{12.8}{\nano\second}). As soon as the stage enable component receives the ready signal from the output ports, it enables the appropriate stage CAM lookup table on the next cycle so that the next packet (whenever it arrives) is forwarded according to the new stage.

Our FPGA prototype demonstrates that it is feasible to implement a fast \LCDC switch with ns-scale latencies. ASIC implementations would be even faster than our FPGA implementation, but they are outside the scope of this paper.

\begin{figure}[t]
    \centering
    \begin{subfigure}{\columnwidth}
       \centering
       \includegraphics[width=\columnwidth]{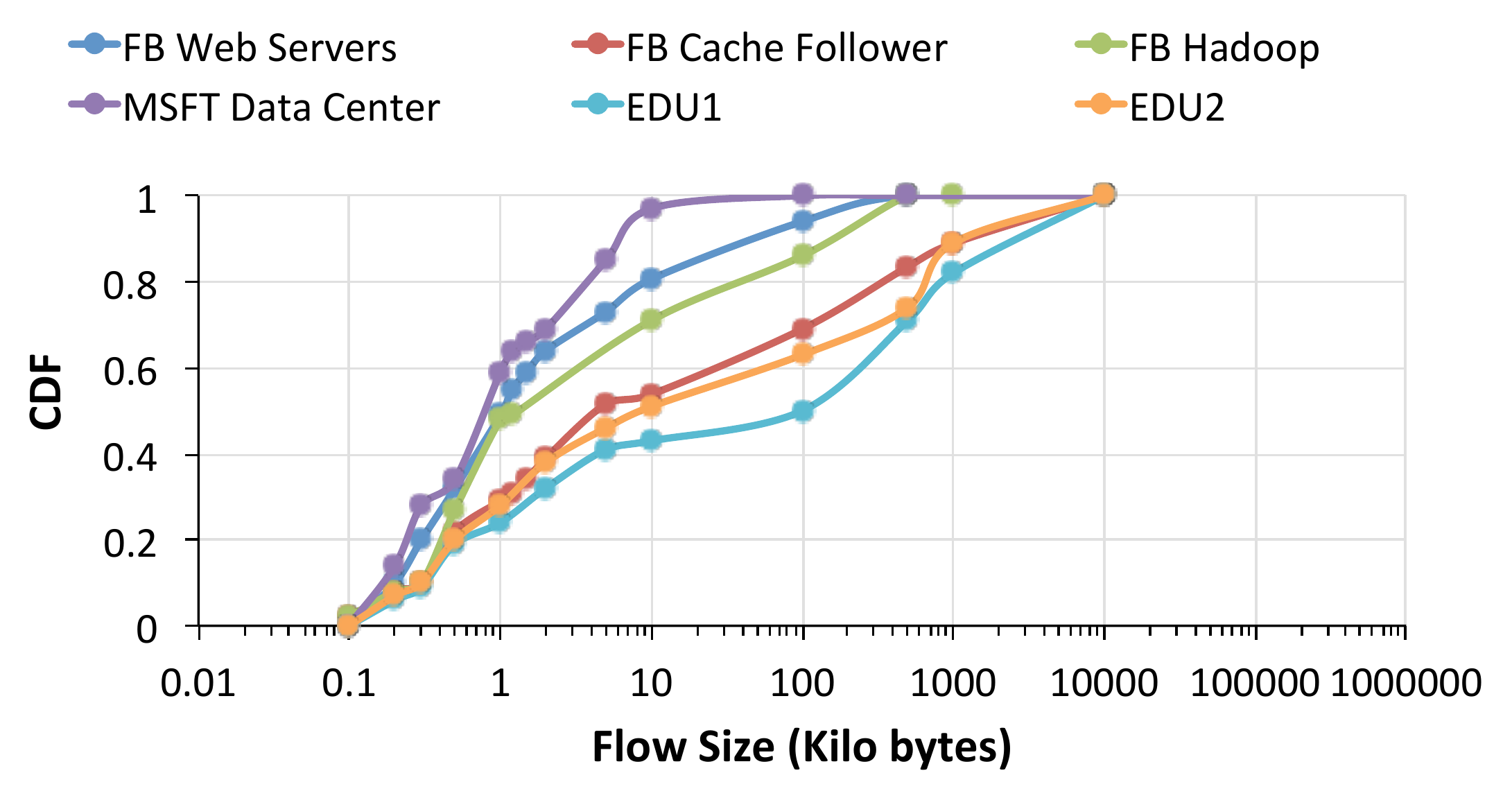} \\[\abovecaptionskip]
       \vspace{-8pt}
       \caption{Flow size CDF.}
       \label{fig:flow_size}
       \vspace{8pt}
    \end{subfigure}
    \hfill
    \begin{subfigure}{\columnwidth}
       \centering
       \includegraphics[width=\columnwidth]{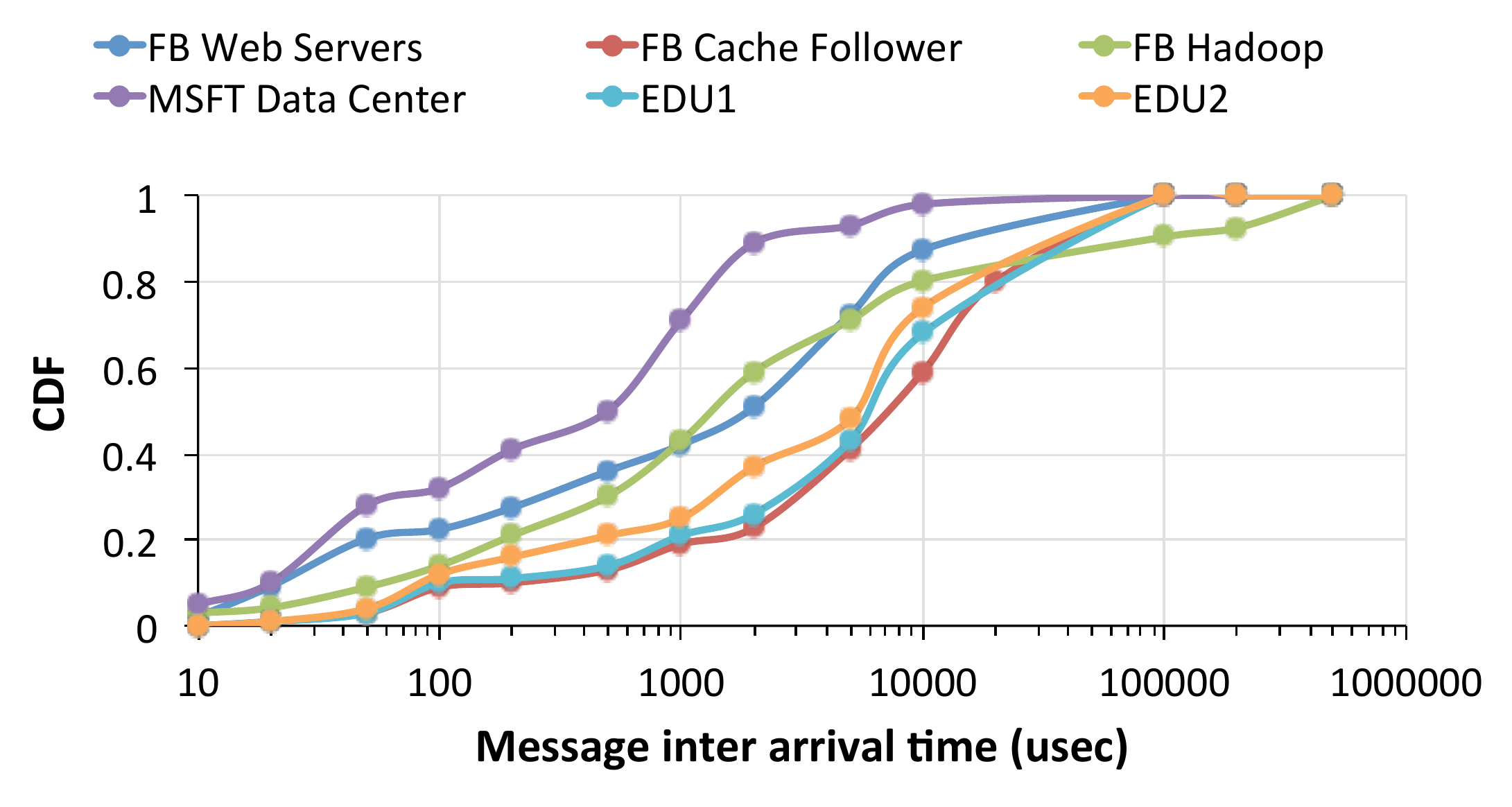} \\[\abovecaptionskip]
       \vspace{-8pt}
       \caption{Flow interval CDF.}
       \label{fig:flow_interval}
    \end{subfigure}

    \caption{\textbf{CDF of traffic data input used in simulation.}}
    \label{fig:data}
\end{figure}

\subsection{Feasibility at the Node Level}
\label{node_feasibility}
Using our modified Linux network device driver we measured the latency between the hypothetical ``laser turn on'' command that the device driver will issue to the NIC transceiver, and the subsequent call to the device-specific transmit function that starts sending the bits through the physical link. To do this we took timestamps with our version of the \texttt{sendmsg()} system call and NIC device transmit function were invoked. Our test-bed consists of an Intel 82579LM Ethernet card and an Intel Core i5 2520M \SI{2.5}{\giga\hertz} processor running Linux kernel 4.2.0 on Ubuntu 15.10. We measured a mean elapsed time of \SI{3.2}{\micro\second} over 100k samples on a completely idle system at runlevel 1 to minimize perturbations from other kernel services, \texttt{TCP{\_}NODELAY} to eliminate small packet aggregation, and with hyper-threading, frequency governors, and all but one cores disabled to minimize perturbations due to thread migration or core frequency changes.

\begin{figure}[t]
   \centering
   \begin{subfigure}{0.49\columnwidth}
      \centering
      \includegraphics[width=\textwidth]{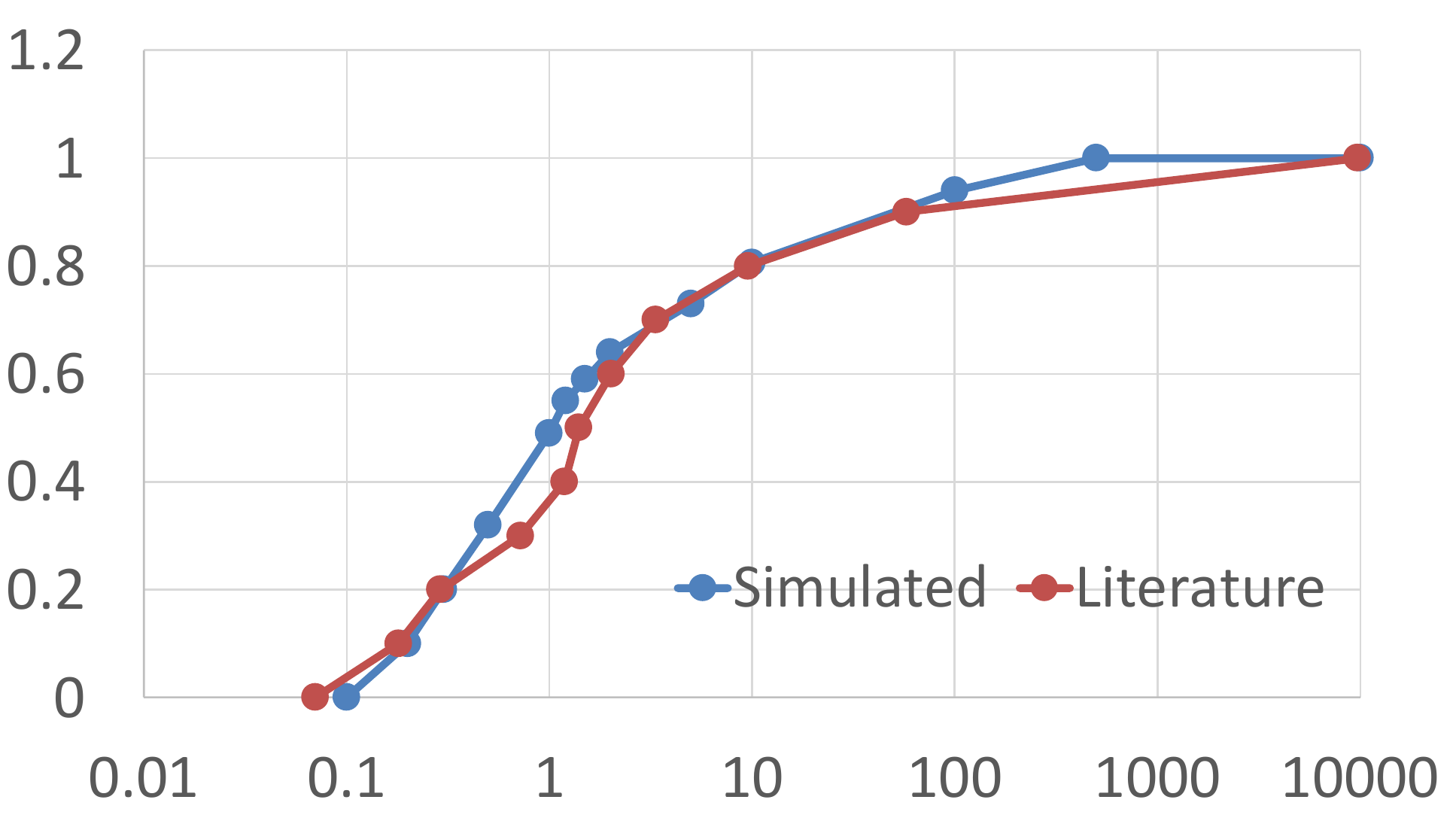} \\[\abovecaptionskip]
      \vspace{-8pt}
      \caption{Flow size: FB web server.}
      \label{fig:comp_flowsize_fb_web_server}
      \vspace{8pt}
   \end{subfigure}
   \hfill
   \begin{subfigure}{0.49\columnwidth}
      \centering
      \includegraphics[width=\textwidth]{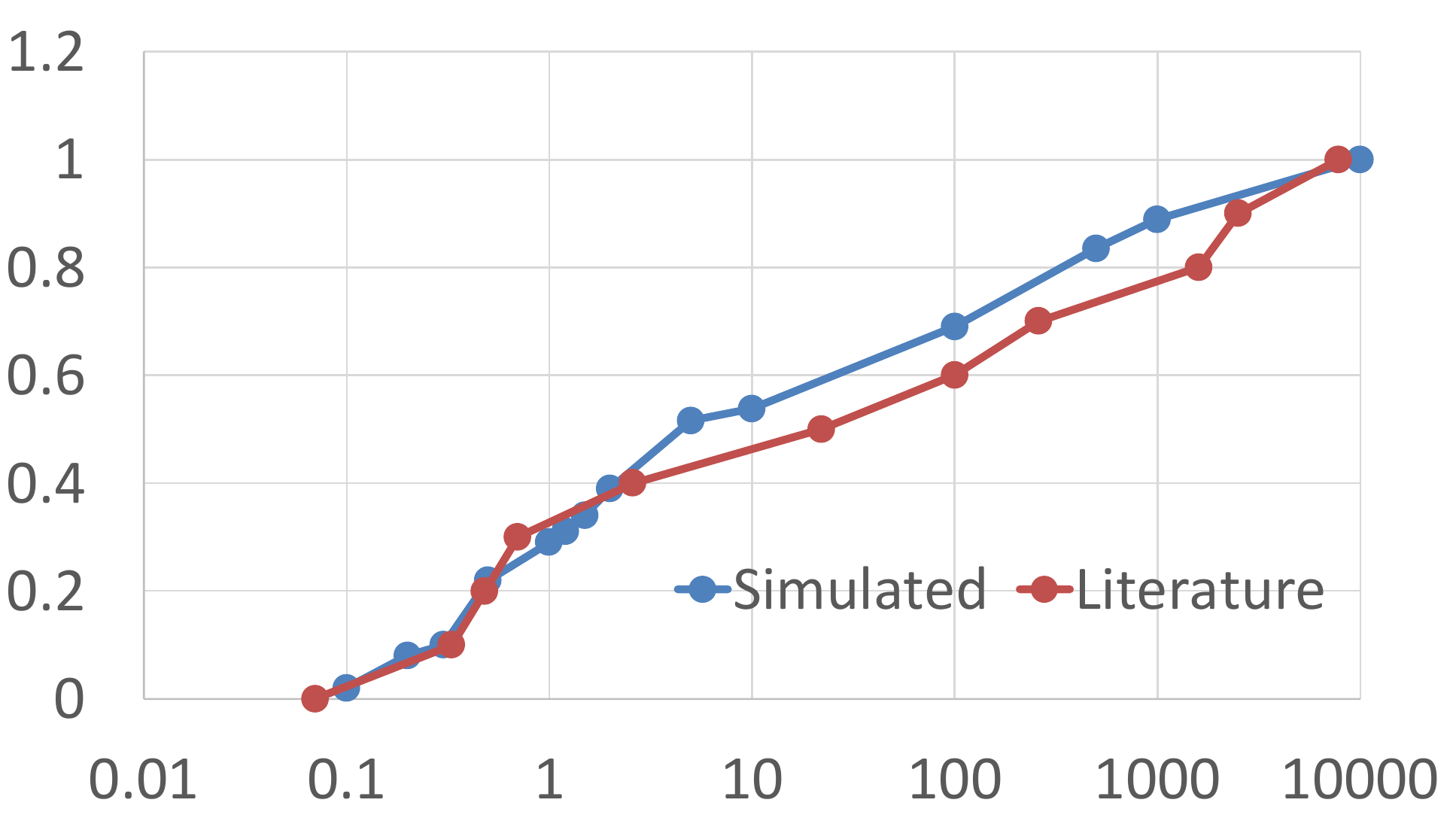} \\[\abovecaptionskip]
      \vspace{-8pt}
      \caption{Flow size: FB cache follower.}
      \label{fig:comp_flowsize_fb_cache_follower}
      \vspace{8pt}
   \end{subfigure}

   \begin{subfigure}{0.49\columnwidth}
       \centering
       \includegraphics[width=\textwidth]{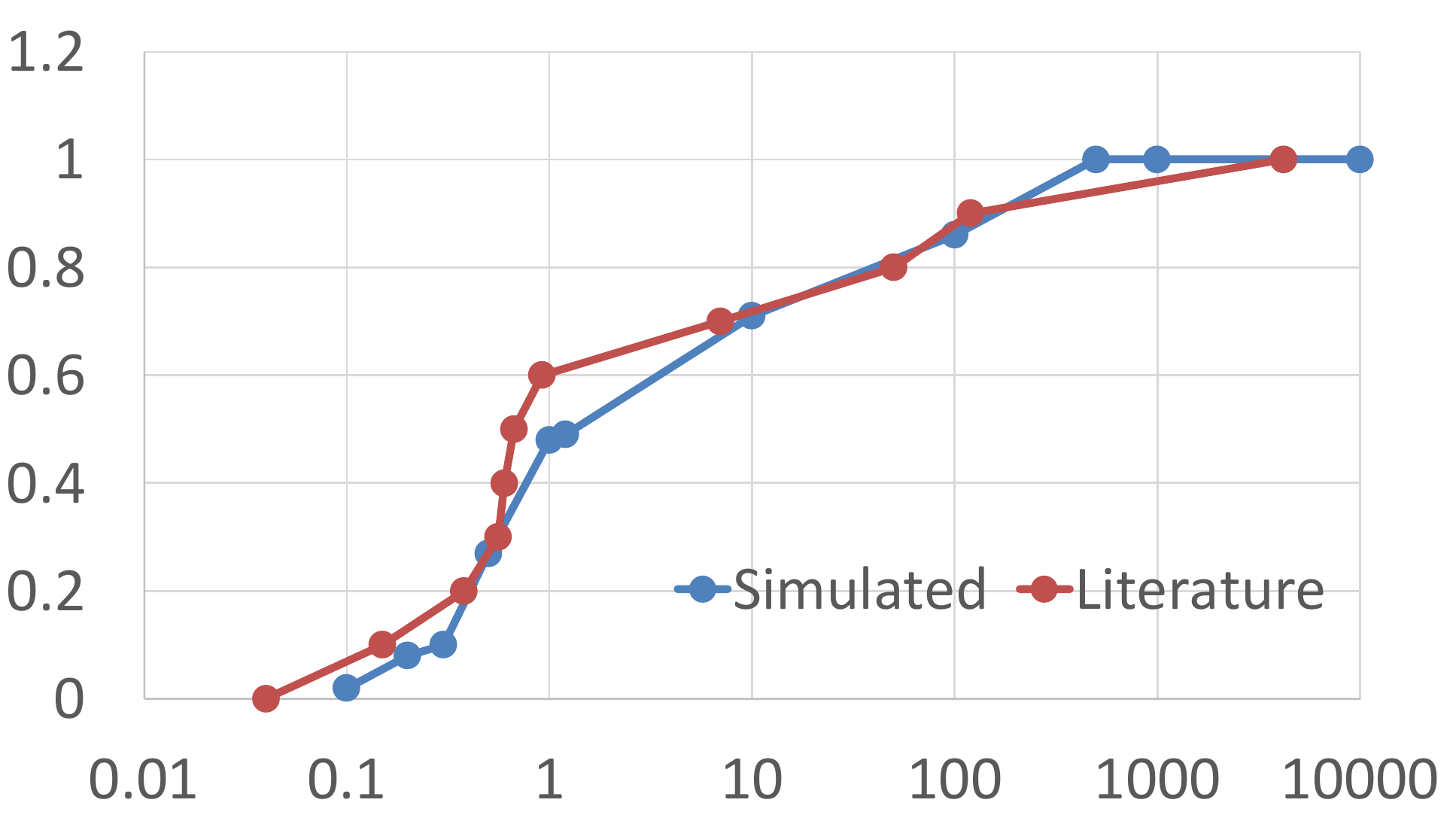} \\[\abovecaptionskip]
       \vspace{-8pt}
       \caption{Flow size: FB Hadoop.}
       \label{fig:comp_flowsize_fb_hadoop}
      \vspace{8pt}
    \end{subfigure}
    \hfill
    \begin{subfigure}{0.49\columnwidth}
       \centering
       \includegraphics[width=\textwidth]{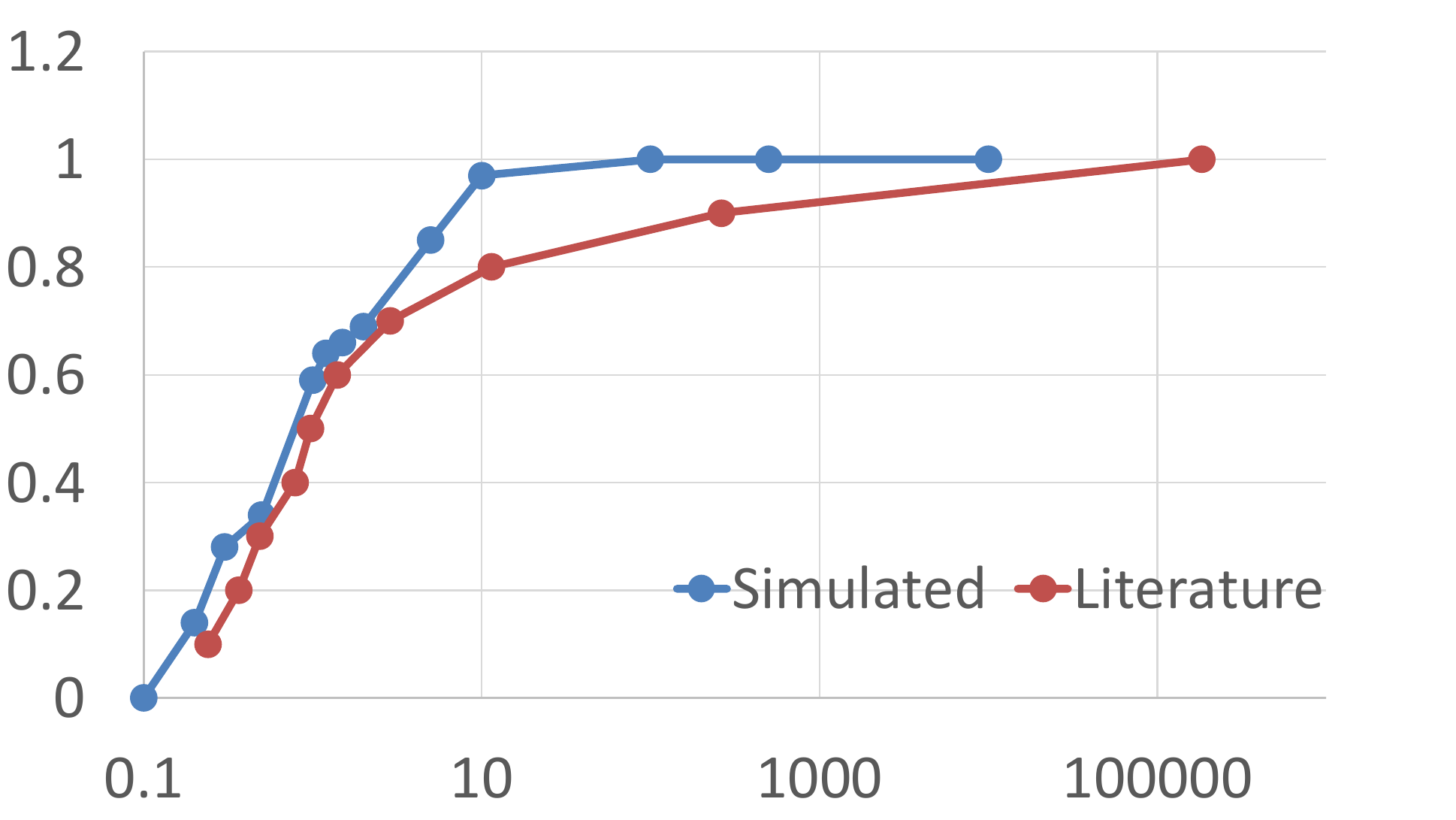} \\[\abovecaptionskip]
       \vspace{-8pt}
       \caption{Flow size: MS Data Center.}
       \label{fig:comp_flowsize_msft}
      \vspace{8pt}
    \end{subfigure}

    \begin{subfigure}{0.49\columnwidth}
       \centering
       \includegraphics[width=\textwidth]{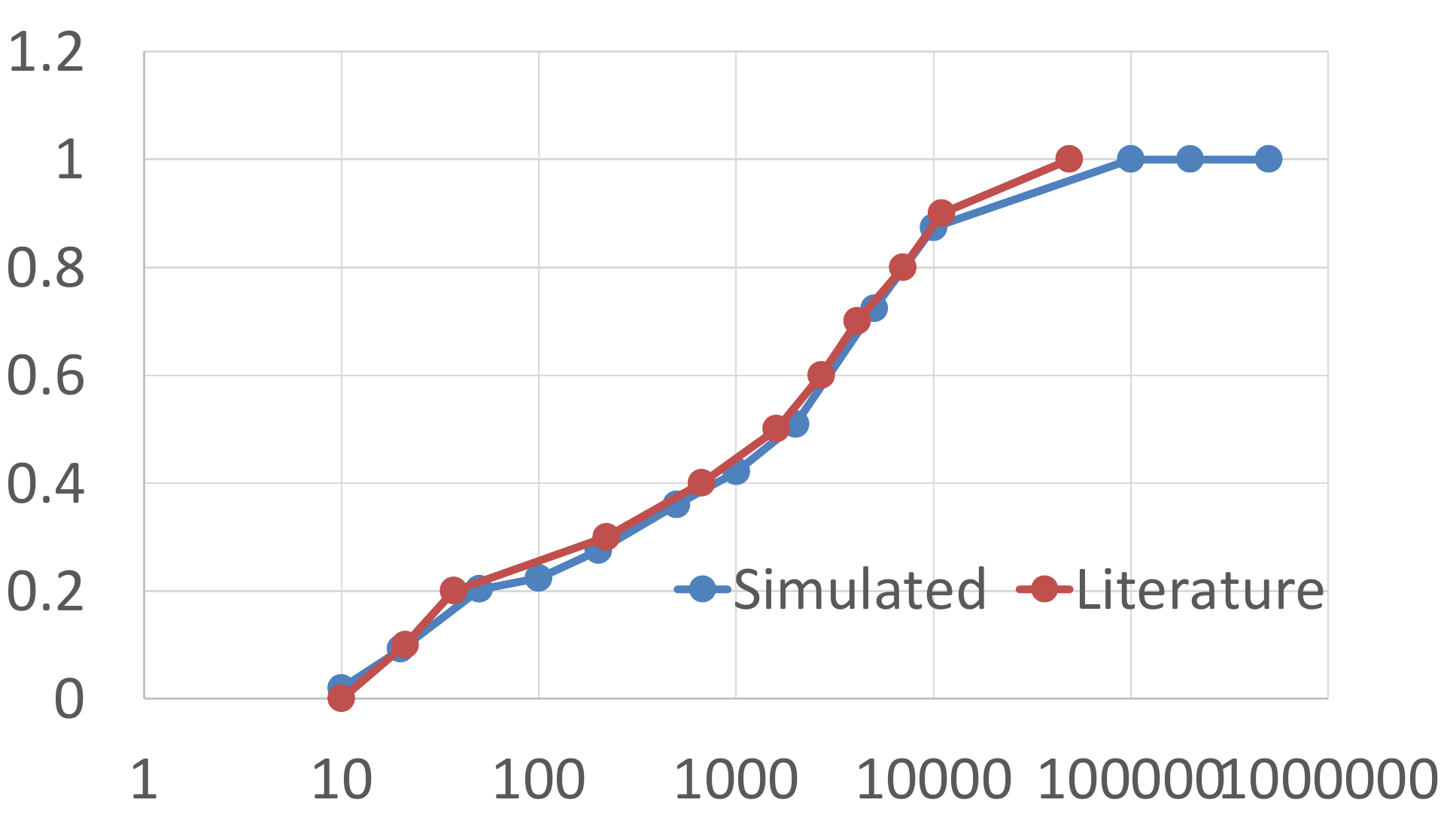} \\[\abovecaptionskip]
       \vspace{-8pt}
       \caption{Flow interval: FB web server.}
       \label{fig:comp_flowinterval_fb_web_server}
      \vspace{8pt}
    \end{subfigure}
    \hfill
    \begin{subfigure}{0.49\columnwidth}
       \centering
       \includegraphics[width=\textwidth]{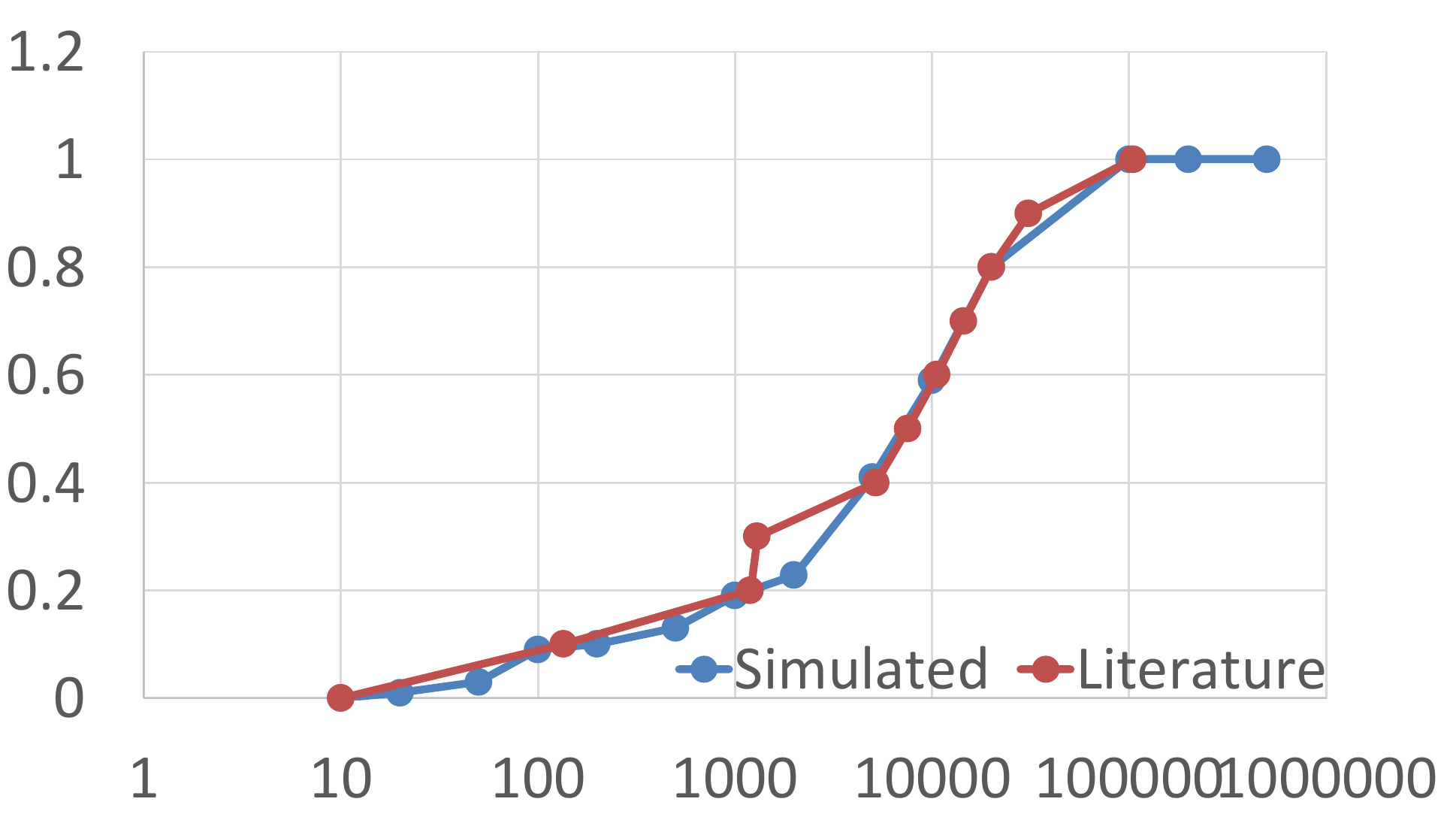} \\[\abovecaptionskip]
      \vspace{-8pt}
       \caption{Flow interval: FB cache follower.}
       \label{fig:comp_flowinterval_fb_cache_follower}
      \vspace{8pt}
    \end{subfigure}

    \begin{subfigure}{0.49\columnwidth}
        \centering
        \includegraphics[width=\textwidth]{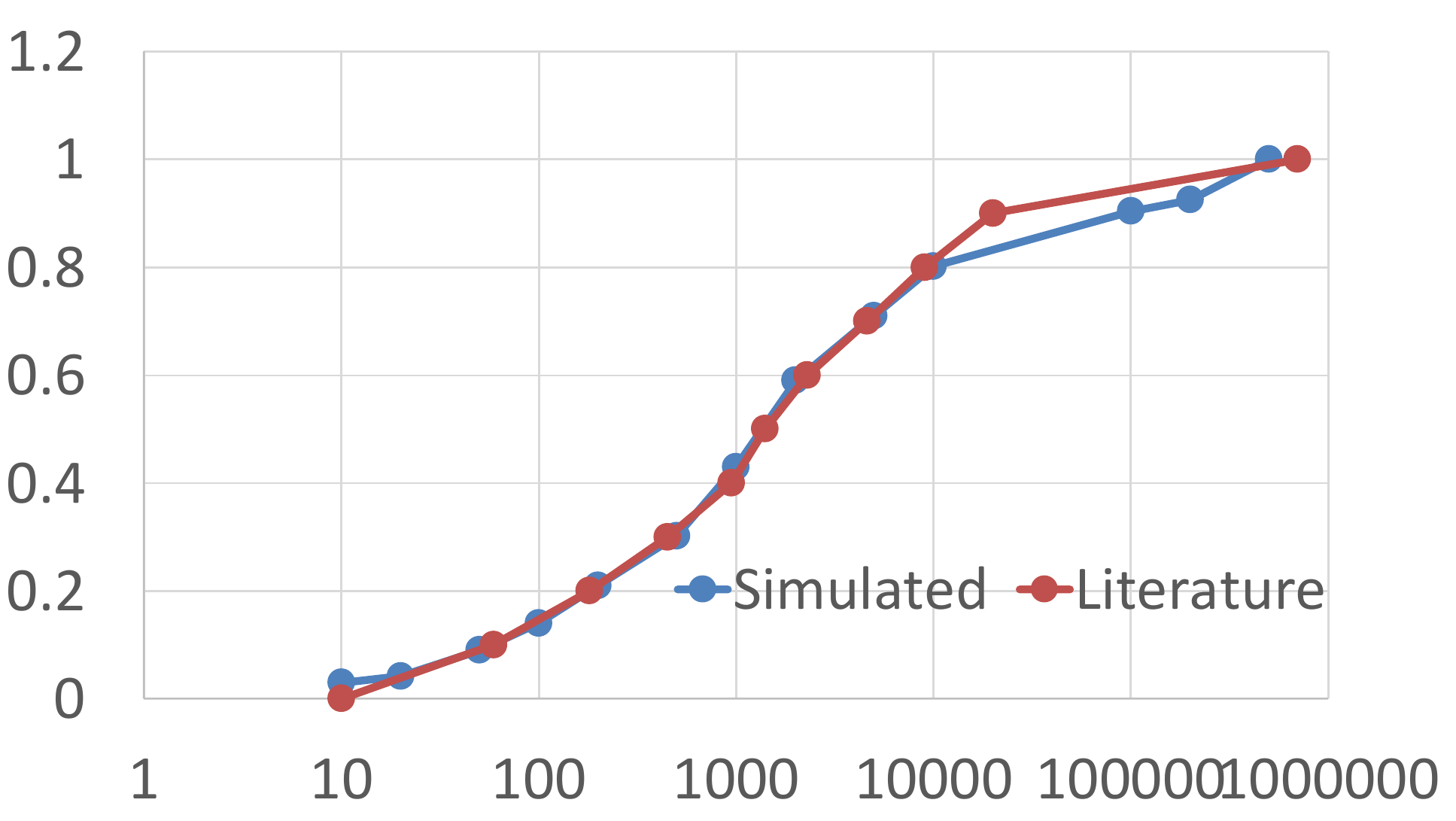} \\[\abovecaptionskip]
        \vspace{-8pt}
        \caption{Flow interval: FB Hadoop.}
        \label{fig:comp_flowinterval_fb_hadoop}
      \vspace{8pt}
     \end{subfigure}
     \hfill
     \begin{subfigure}{0.49\columnwidth}
        \centering
        \includegraphics[width=\textwidth]{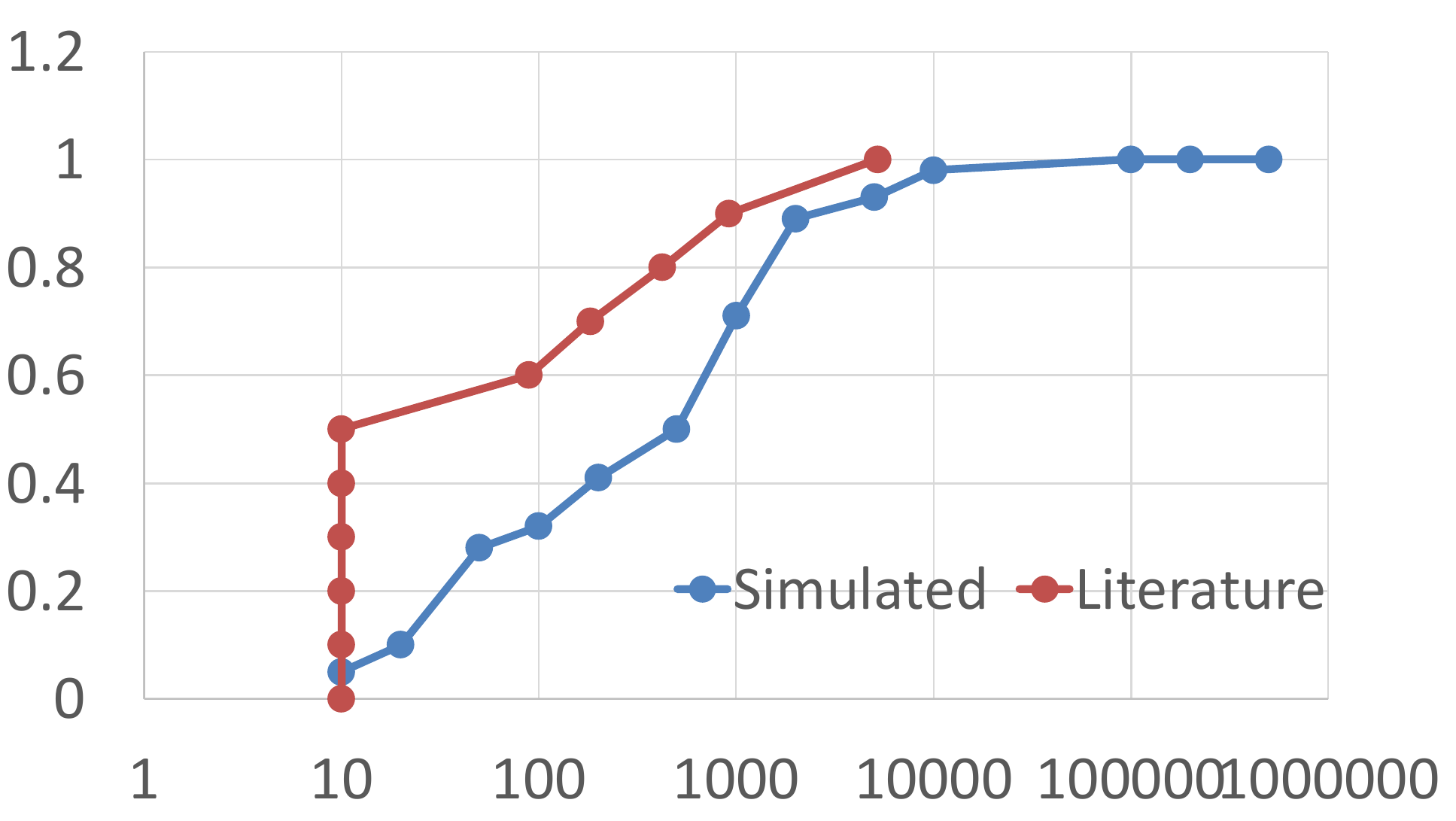} \\[\abovecaptionskip]
        \vspace{-8pt}
        \caption{Flow interval: MSFT Data Center}
        \label{fig:comp_flowinterval_msft}
      \vspace{8pt}
     \end{subfigure}

   \caption{\textbf{Comparison of simulated traffic CDF of flow size and flow intervals vs. target large-scale data center traffic from published measurements.}}
   \label{fig:comp}
\end{figure}

Our results corroborate independent measurements in literature of \SI{3.7}{\micro\second} for a packet to traverse the TCP/IP stack~\cite{larsen:tcplatency}. These results independently measured that it
takes \SI{950}{\nano\second} for a process to send a message to the socket
interface on a connection that has already been established.
Evoking a socket write begins the TCP layer to initiate
transmission, copy the application buffer into the transmit
queue in kernel space and prepare a datagram for the IP layer
(\SI{260}{\nano\second}). Then the IP layer does routing, segmentation,
processes the IP header, and eventually calls the network
device driver (\SI{550}{\nano\second}). The network device driver constructs
the output packet queue entry and calls the precise hardware
implementation of the NIC card to transmit the frame by
passing a pointer to the packet descriptor (\SI{430}{\nano\second}). This
causes a control register write within the NIC to set up a
DMA transfer to fetch the pointer, and when it completes
control is handed to the NIC card (\SI{400}{\nano\second}).
Another \SI{760}{\nano\second}
are consumed by the NIC to process the core register write,
interpret the descriptor, and based on the descriptor initiate a
DMA to fetch from main memory the data of the packet to
transmit. Each 64-byte cache line access to memory takes an
estimated \SI{400}{\nano\second} to propagate from the PCIe signal pins to
memory and back. Thus, it takes a total of \SI{3.75}{\micro\second} for an
application to launch a packet onto the fiber interface.
Thus, the server's NIC card will have ample time to turn on its laser and it will impose no laser turn-on delay to the sending process.

%% file: methodology.tex
\section{Experimental Methodology}
\label{methodology}

\begin{figure}[t]
   \centering
   \includegraphics[width=\columnwidth]{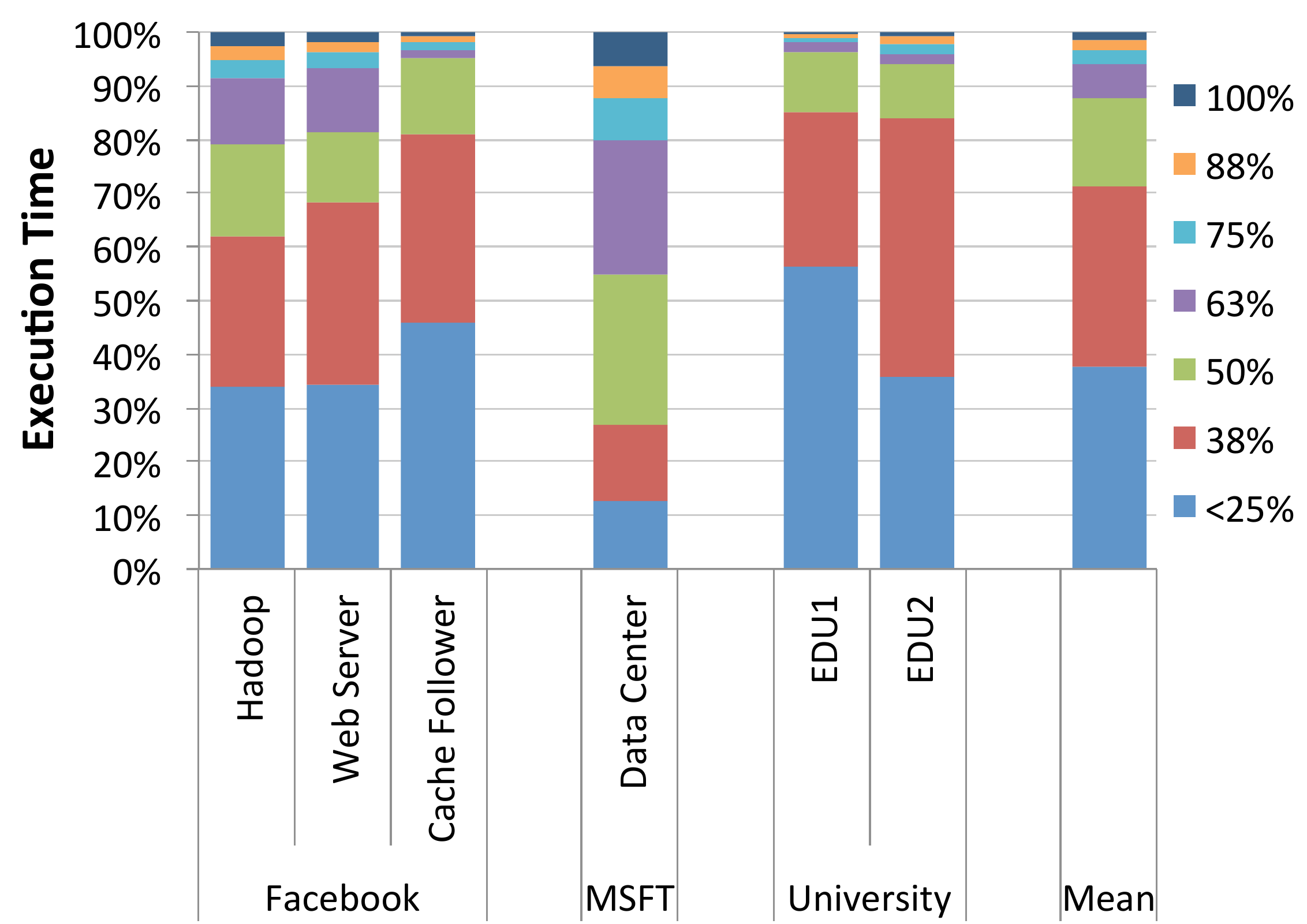}
   \caption{\textbf{Breakdown of partial network activation. The legend indicates the fraction of the network links that are on.}}
   \label{fig:activation}
\end{figure}

\begin{figure}[t]
   \centering
   \includegraphics[width=0.96\columnwidth]{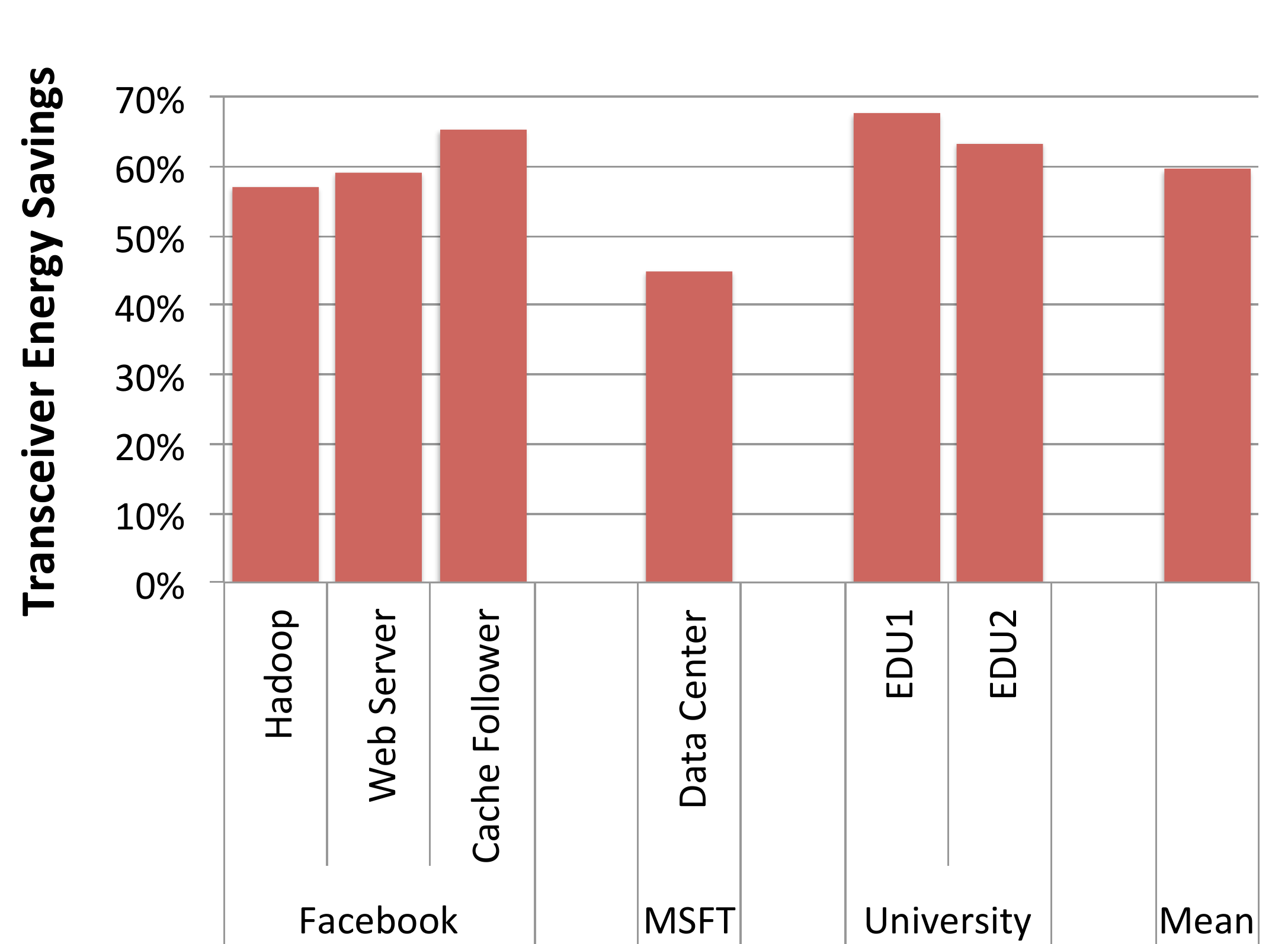}
   \caption{\textbf{\LCDC transceiver energy savings.}}
   \label{fig:TxEnergy}
\end{figure}

\begin{figure}[t]
   \centering
   \includegraphics[width=0.96\columnwidth]{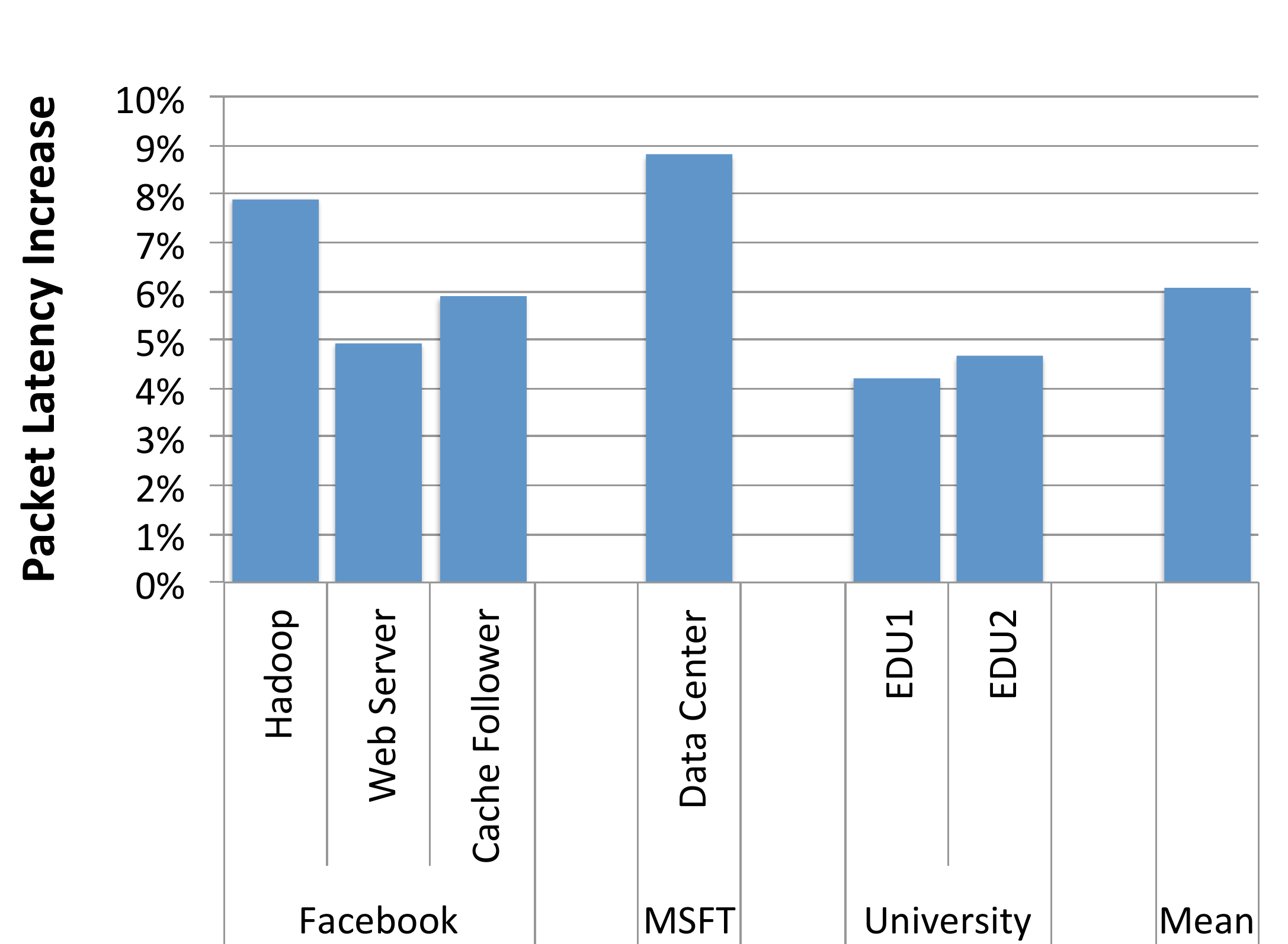}
   \caption{\textbf{Impact of \LCDC on packet latency.}}
   \label{fig:latency}
\end{figure}

\begin{figure*}[t]
   \begin{subfigure}{0.32\textwidth}
       \centering
       \includegraphics[width=\textwidth]{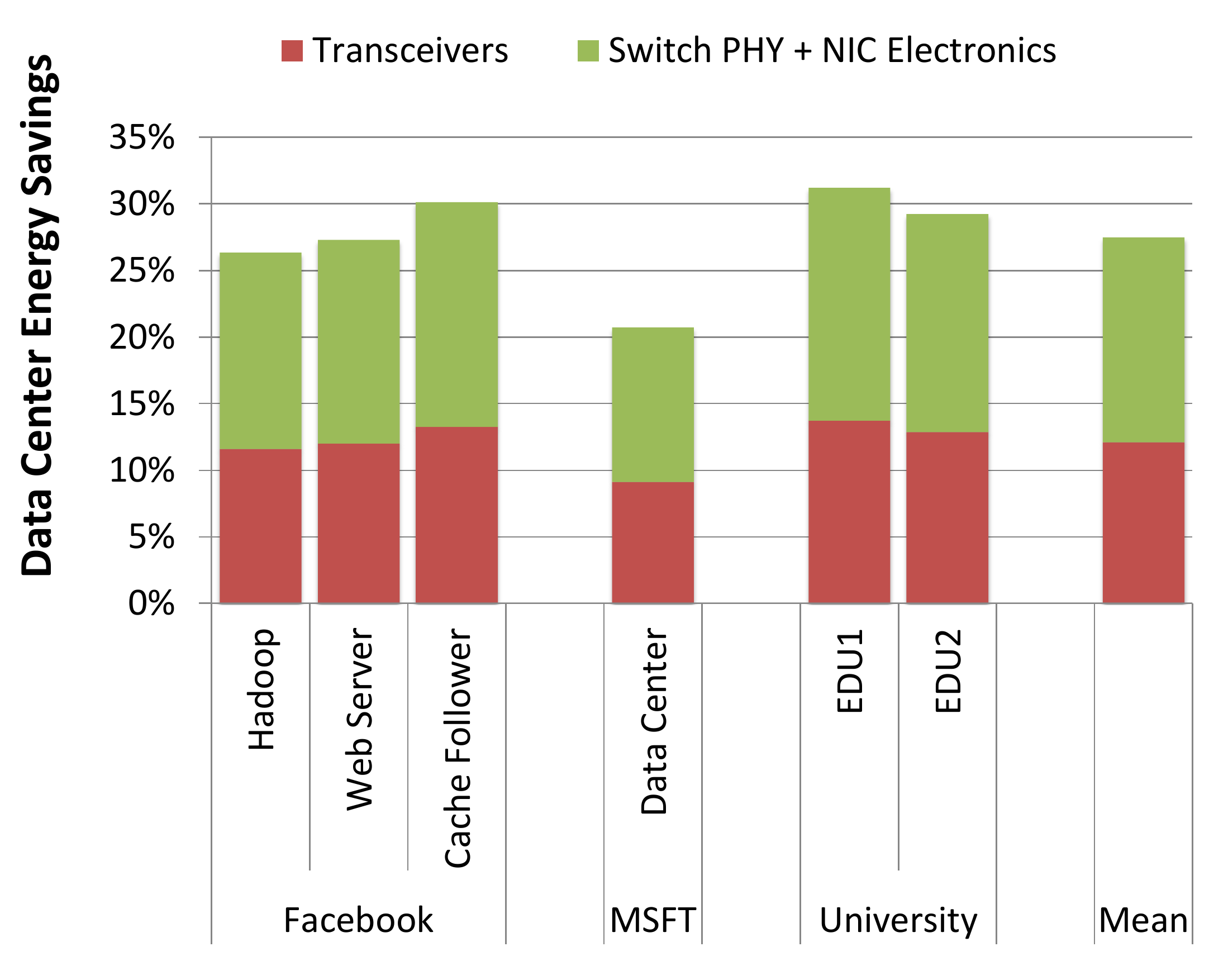} \\[\abovecaptionskip]
       \caption{Cloud servers at 30\% utilization.}
       \label{fig:overall-30}
    \end{subfigure}
    \begin{subfigure}{0.32\textwidth}
       \centering
       \includegraphics[width=\textwidth]{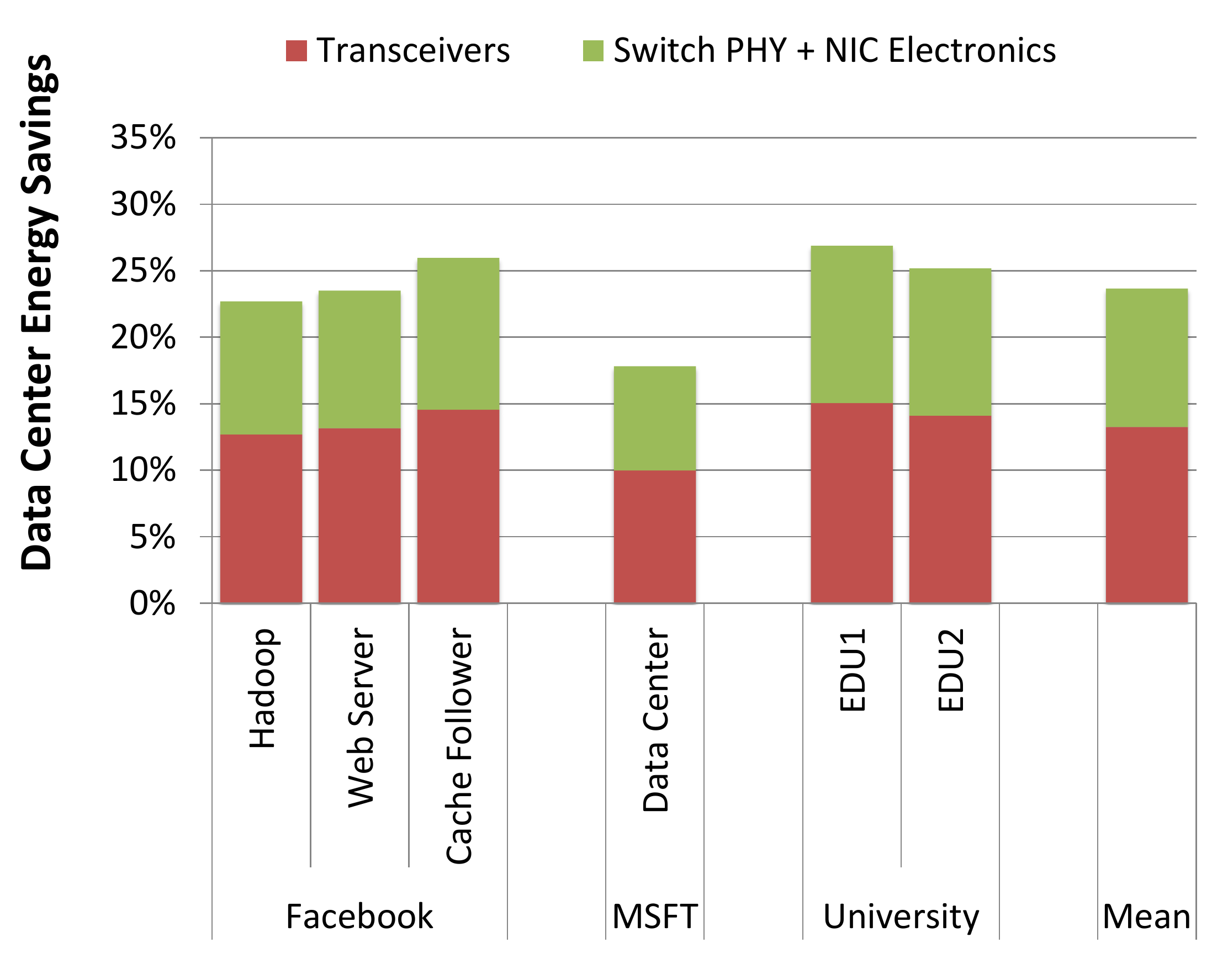} \\[\abovecaptionskip]
       \caption{Cloud servers at 50\% utilization.}
       \label{fig:overall-50}
    \end{subfigure}
    \begin{subfigure}{0.32\textwidth}
       \centering
       \includegraphics[width=\textwidth]{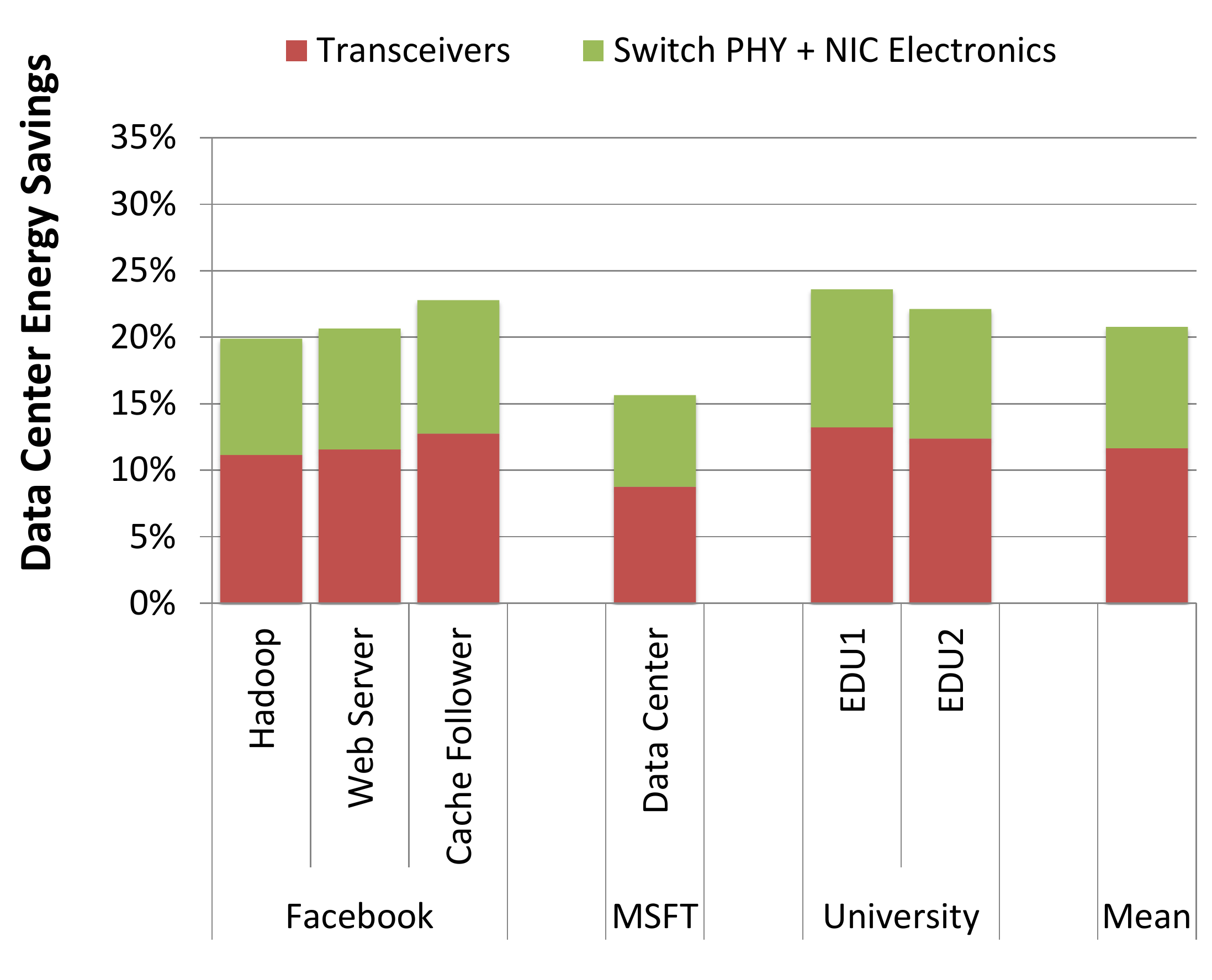} \\[\abovecaptionskip]
       \caption{Cloud servers at 70\% utilization.}
       \label{fig:overall-70}
    \end{subfigure}
   \caption{\textbf{Impact of \LCDC on overall data center energy.}}
   \label{fig:overall}
\end{figure*}

We model the data center network in Figure~\ref{fig:config} using the BookSim cycle-accurate network simulator~\cite{dally:NoCbook}
that we modified in-house.
We faithfully model all traffic, including the additional control packets generated by \LCDC.
The simulator is supplied packets from a traffic generator that shapes traffic to conform to the distributions
prevalent in large-scale data centers. The traffic generator models the traffic characteristics shown in~\cite{greenberg:vl2, kandula:traffic-nature, roy:facebook-network}.
In addition to large-scale data centers with heavy traffic, we evaluate the performance of \LCDC on more traditional data centers
that exhibit lower traffic demand. For that, we use snippets of traces collected from routers in a university data center~\cite{benson:edutrace}.

The feasibility study (Section~\ref{feasibility}) demonstrated that optical devices can turn on/off at \si{\nano\second}-\si{\micro\second} scale.
To remain conservative in our evaluation, we model laser turn on/off times based on a commercially available SFP+ module
(MRV SFPFC401) which has a turn-on/off delay of \SI{1}{\micro\second} /\SI{10}{\micro\second}~\cite{mrv:SFPFC401}.
We model the delay characteristics of routers as derived from our feasibility study (Section~\ref{feasibility}),
and we calculate link latency based on the traversed optical fiber length.

We estimate network performance by measuring the average packet delivery latency, and the energy savings by measuring the fraction of the time each link is deactivated. We set the high watermark at 75\% buffer utilization for stage activation, and the low watermark at 22\% buffer utilization for stage deactivation (experimentally determined to balance energy savings with network performance).

%% file: results.tex
\section{Experimental Results}
\label{results}

\subsection{Input Data}
\label{data}

Figure~\ref{fig:data} shows the data traffic injected into our simulated network. We created a traffic generator that produces traffic that closely approximates the network traffic originating from large-scale data centers in Facebook~\cite{roy:facebook-network}, Microsoft~\cite{greenberg:vl2, kandula:traffic-nature}, as well as in higher education settings~\cite{benson:edutrace}. Comparing the distribution of traffic from our generator with the large-scale data center traffic from published measurements~\cite{roy:facebook-network, greenberg:vl2, kandula:traffic-nature}, we confirm that they have similar CDFs, as is shown in Figure~\ref{fig:comp}. We measure the Pearson r coefficient to be between 0.979--0.992 for the flow size CDF, and 0.894--0.998 for the flow interval CDF.
Thus, our simulations are conducted under conditions that closely approximate real-world environments.

\subsection{Results}
\label{resultsub}

Figure~\ref{fig:activation} shows the portion of the network that is activated during the execution of the traffic workload for each modelled traffic. Most traffic types exhibit sparse and bursty packet injection trends. Thus, \LCDC finds windows of low utilization to deactivate a link, and 87\% of the time on average half of the network is deactivated, indicating the potential to achieve significant power savings. The Microsoft data center traffic presents the most challenges, but \LCDC still manages to turn off half of the network half the time.

As a result, \LCDC saves on average 60\% of the optical transceivers' energy (Figure~\ref{fig:TxEnergy}). The lower energy savings relative to the time breakdown (Figure~\ref{fig:activation}) are due to the fact that while a transceiver is in the process of turning on or off, the link is still considered deactivated, but we conservatively charge the full transceiver power consumption to the network. The energy savings come at the cost of 6\% higher average packet latency (Figure~\ref{fig:latency}) as queueing in network buffers may slightly increase, which we argue can be largely absorbed by the application layer.

Following the analysis in Section~\ref{motivation} we estimate the data center energy savings of \LCDC in Figure~\ref{fig:overall}. We consider the data center to be at an average utilization of 30\%. Additionally we calculate the energy savings of \LCDC in cloud servers, which are shown to typically have higher utilizations at 40\%-70\%~\cite{nrdc:efficiency}. Figures~\ref{fig:overall-50} and~\ref{fig:overall-70} show energy savings at 50\% and 70\% utilization, respectively.
For data centers at 30\% utilization, assuming the data center servers are optimized for energy while maintaining high performance, \LCDC can save 12\% of the data center energy by deactivating links when they are not needed. It is also possible to extend \LCDC to deactivate or put into a sleep mode the switch PHY chips and the server NIC electronics whenever the corresponding link is deactivated. With that extension, \LCDC's data center energy savings can reach up to 27\% on average, even after accounting for the CMOS scaling of the switch PHY and NIC card electronics. While we do not directly explore turning off the switch PHY and NIC card electronics, it is a promising future direction for this work.
For cloud servers with utilizations of 50\% and 70\%, \LCDC can respectively save 13\% and 12\% of total data center energy by just turning off links. By deactivating the switch PHY and NIC electronics, \LCDC can respectively save 23\% and 21\% of total data center energy.

%% file: related.tex
\section{Related Work}
\label{related}
Heller \textit{et. al} propose ElasticTree~\cite{heller:elastic-tree}, a power manager that dynamically adjusts the set of active links and switches to satisfy data center traffic while saving energy. However, ElasticTree cannot hide the high latency of rebooting a switch, and thus it is only applicable to large-grain traffic ebb and flow, while \LCDC exerts very fine-grain control over link activations.
Ananthanarayanan and Katz~\cite{ananthanarayanan:greening-switch} present a switch design that estimates traffic to power down ports when possible. Unlike \LCDC, it may send packets to sleeping ports and expose long delays to the application.

Laser gating has been proposed before for on-chip interconnects~\cite{demir:ecolaser, demir:ecolaser-plus, demir:lac, demir:slac}, which however do not present the complexities that laser gating on data center networks has to address (e.g., fast CDR, fast power-up of commercial transceiver electronics, control plane updates at the switch, and kernel modification to set up an early warning system). A previous proposal for laser control in on-chip interconnects~\cite{demir:slac} was extended to data centers but only as a conceptual study; no evaluation was performed on traffic similar to modern large-scale data centers (only a university data center traffic trace was used), and the feasibility of the technique was not demonstrated.

Helios~\cite{farrington:helios} identifies the subset of traffic best suited to circuit switching and dynamically reconfigures the network topology at runtime based on shifting communication patterns. Unlike \LCDC, Helios exhibits a high reconfiguration cost of \SI{15}{\milli\second}, even with Electronic Dispersion Compensation disabled (noise compensation is not needed when light from high-power transceivers travels over a short fiber).
Helios needed to constantly reconfigure the topology, but did not capitalize on fast CDR techniques such as clock phase caching nor considered transceivers that are built to support fast on/off timing. It relied on commercially-available components that were designed over a decade ago at a time when there was no incentive to make control plane updates fast, as traditionally they were infrequent events; updates on many such switches are happening by posting a request to perform an update, which sometime later is picked up by a polling thread within the switch and is serviced. In contrast, \LCDC is designed from the ground up with speed as one of the main targets, and the software and hardware infrastructure are redesigned around it.

%% file: conclusion.tex
\section{Conclusion}
\label{conclusion}
As technological innovations steadily reduce the power consumption of data center components, network power consumption becomes increasingly prominent. We argue that it is time to start optimizing the network designs not only for latency, bandwidth and cost, but also for power and energy efficiency. We present \LCDC, a data center network system architecture in which the operating system, the switch, and the optical components are co-designed to achieve energy proportionality. We demonstrate \LCDC's feasibility at all levels (electrical circuitry, optical devices, node-level architecture, and switch architecture) and show that it can save the majority of the network energy in a data center.